\documentclass[useAMS,usenatbib]{mn2e}

\usepackage{graphicx}

\usepackage{natbib}
\usepackage{color}
\newcommand{\feh}{\mbox{[Fe/H]}}
\newcommand{\afe}{\mbox{[$\alpha$/Fe]}}
\newcommand{\cfe}{\mbox{[C/Fe]}}
\newcommand{\nfe}{\mbox{[N/Fe]}}
\newcommand{\ofe}{\mbox{[O/Fe]}}
\newcommand{\nafe}{\mbox{[Na/Fe]}}

\newcommand{\tife}{\mbox{[Ti/Fe]}}
 \newcommand{\sife}{\mbox{[Si/Fe]}}
\newcommand{\cafe}{\mbox{[Ca/Fe]}}
\newcommand{\crfe}{\mbox{[Cr/Fe]}}

\newcommand{\mgfe}{\mbox{[Mg/Fe]\arcmin}}

\def\akn{}
\def\akn{\color{black}}

\title[Globular clusters in KKs\,3 and ESO\,269-66]{Nuclei of dwarf spheroidal galaxies KKs\,3 and 
ESO\,269-66 and their counterparts in our Galaxy}
\author[M. E. Sharina et al.]{M. E. Sharina$^{1}$,
V. V. Shimansky$^{2}$, and A. Y. Kniazev$^{\akn 3,4,5,1}$ \\
$^{1}$Special Astrophysical Observatory, Russian Academy of
Sciences, N. Arkhyz, KCh R, 369167, Russia\\
$^{2}$Kazan Federal University, Kremlevskaya 18, Kazan, 420008, Russia\\
$^{3}$South African Astronomical Observatory, PO Box 9, 7935 Observatory, Cape Town, South Africa \\
$^{4}$Southern African Large Telescope Foundation, PO Box 9, 7935 Observatory, Cape Town, South Africa \\
$^{5}$Sternberg Astronomical Institute, Lomonosov Moscow State University, Moscow, Russia\\
}

\begin{document}

\date{Accepted . Received ; in original form }

\maketitle

\begin{abstract}
We present the analysis of medium-resolution spectra obtained at the Southern African Large
Telescope (SALT) for nuclear globular clusters (GCs) in two dwarf spheroidal galaxies (dSphs).
The galaxies have similar star formation histories, but they are situated in completely
different environments. ESO\,269-66 is a close neighbour of the giant S0 NGC\,5128. 
  KKs\,3 is one of the few truly isolated dSphs within 10 Mpc.
We estimate the helium abundance $Y=0.3$, $\rm age=12.6\pm1$~Gyr, $\feh=-1.5,-1.55\pm0.2$~dex,
and abundances of C, N, Mg, Ca, Ti, and Cr for the nuclei of ESO\,269-66 and KKs\,3.
Our surface photometry results using HST images yield the half-light radius of the cluster in KKs\,3, $\rm r_h=4.8\pm0.2$~pc.
We demonstrate  the similarities of medium-resolution spectra, ages, chemical compositions, and structure
for GCs in ESO\,269-66 and KKs\,3 and for several massive Galactic GCs with $\feh\sim-1.6$~dex. 
All Galactic GCs posses Extended Blue Horizontal Branches and multiple stellar populations. 
Five of the selected Galactic objects are iron-complex GCs.
Our results indicate that the sample GCs observed now
in different environments had similar conditions of their formation $\sim$1 Gyr 
after the Big Bang.

\end{abstract}


\section{Introduction}

Many of the early-type dwarf galaxies contain old massive ($10^6\div10^7 M_{\odot}$) nuclei
near their optical centres 
(Lotz et al. 2004, Sharina et al. 2005; C{\^o}t{\'e} et al. 2006, Georgiev et al. 2009, 
Da Costa et al. 2009). These are located at the extreme high end of
the GC mass function and constitute a large fraction of the total mass of these
galaxies \citep{2012A&A...546A..53L}.
The mass of the nuclear star cluster correlates with the mass of the stellar
spheroids of their host galaxies (Leigh et al. 2012).
Characteristics of the nuclei provide important information about the physical
conditions of the early host galaxy formation and evolution.

The existence of the phenomenon of multiple stellar populations in GCs
is not completely understood (Charbonnel 2016 and references therein).
One of the ideas is that these GCs are remnant nuclei of dwarf galaxies. 
In recent years, the number of observational facts related to this phenomenon
 has greatly increased (Gratton et al. 2012, Charbonnel 2016).
 High-resolution spectroscopic observations revealed correlated variations of C, N, O, Na, Al, Mg 
abundances for stars in almost all old Galactic GCs (e.g. Carretta et al. 2010 and references therein).
Chemical patterns of many Galactic GCs indicate that they contain 
several generations of stars and that the material
from which GCs were formed was exposed to proton-capture processes at high temperatures.
First stellar generations have chemical composition similar to that of the 
field stars. Second and subsequent (if any) stellar populations can be depleted in C and O,
enhanced in N and Na.
Helium enrichment in GCs appears to be correlated with the appearance of Extended Blue Horizontal Branches 
(EHBs) and of Na--O and Al--Mg anti-correlations (D'Antona et al. 2002, Salaris et al. 2006,
Chantereau et al. 2016). 
Hubble Space telescope (HST) observations in photometric filters sensitive to C, N, and O abundance 
variations allowed to disentangle splits of the main evolutionary sequences of Galactic 
and nearby extragalactic GCs (e.g. Piotto et al. 2015).
A rare class of anomalous Galactic GCs was discovered 
(Table 10 in Marino et al. 2015 and references therein). 
These objects show variations in Fe and s-process element abundances.
The most plausible explanation for the origin of these iron complex GCs 
is that they are remnant nuclei of dwarf galaxies (Bekki \& Norris 2006).

Properties of our sample dSphs and their nuclei are summarized in
Tables~\ref{tab:prop_gal} and ~\ref{tab:prop_gcs}.
ESO\,269-66 is a dSph satellite of the peculiar giant elliptical Centaurus~A$=$NGC\,5128.
The Cen~A group with its numerous dwarf galaxies and tidal substructures is the only 
elliptical-dominated group within 10~Mpc (Karachentsev et al. 2007, Crnojevi{\'c} et al. 2016).
The projected separation between NGC\,5128 and ESO\,269-66 is $\sim$198~kpc, which is
typical of dSphs (Karachentsev et al. 2005, 2013).
A considerable population of asymptotic giant branch (AGB) stars as 
a signature of the 2-3 Gyrs old star formation was detected in ESO\,269-66 (Crnojevi{\'c} et al. 2011). 
Stellar photometry using Hubble Space Telescope (HST) optical and infrared images indicates
that at least three powerful star forming bursts happened in the galaxy
2$\pm$1.5, 3$\pm$1.5 (Crnojevi{\'c} et al. 2011), and 12-14 Gyr ago (Makarova et al. 2007).
The last of the mentioned events was the most powerful.
The mean metallicity of the oldest red giant branch (RGB) stars in ESO\,269-66 is $\feh\!\sim-1.75$~dex 
(Makarova et al. 2007).
The stellar metallicity dispersion in ESO\,269-66 is surprisingly large for a faint dSph 
(Karachentsev et al. 2007, Sharina et al. 2008).
The dispersion of the stellar metallicities in ESO\,269-66 is comparable to that
of NGC\,5237, a compact I0 type object with a central starburst,
a suggested galaxy collision remnant (Thompson, 1992).

\begin{table}
\caption{Properties of dSphs KKs\,3 and ESO\,269-66.} 
\label{tab:prop_gal}
\begin{center}
\begin{tabular}{lrr}
\hline\hline
                &  KKs\,3 &  ESO\,269-66           \\
\hline
RA(J200.0)      & $\rm 2^h24^m44\fs4$ $^a$& $\rm 13^h13^m09\fs1$ $^b$\\
Dec.(J200.0)     & $\rm -73\degr30\arcmin51\arcsec$ $^a$& $\rm -44\degr53\arcmin24\arcsec$ $^b$\\
$\rm E(B-V)$     & 0.045$^a$ & 0.093 $^b$\\ 
Distance, Mpc   & 2.12 $^a$& 3.82 $^b$   \\
Diameter, kpc   & 1.5 $^a$ & 2.4 $^b$  \\
$\rm V_T$, mag      & 14.47 $^c$& 13.74 $^e$  \\
$\rm (V-I)$, mag  & 0.77 $^c$ & 1.06 $^e$  \\
$\rm M_V$           & -12.3 $^c$& -14.4 $^b$   \\
$\rm M_{HI}/L_B$    & 0.03 $^c$& $<0.002$ $^b$  \\
 $\rm M_{HI}$ , $M_{\sun}$                  & 1.1$\cdot 10^{5}$ $^c$& $<0.9\cdot 10^{5}$~$^b$\\
$\rm SFR_{12 \div14Gyr}$, $M_{\sun} yr^{-1}$& 8.7$\cdot 10^{-3}$ $^a$& 11$\cdot 10^{-3}$~$^d$ \\
$\feh_{12 \div14Gyr}$,dex               & -1.9 $^a$ & -1.75 $^d$  \\
\hline
\end{tabular}
\end{center}
\medskip
Notes: $^a$: Karachentsev et al. (2015b), $^ b$: Karachentsev et al. (2013), 
$^c$:~Karachentsev, Kniazev \& Sharina (2015a), $^d$: Makarova et al. (2007), $^e$ Sharina et al. (2008).
\end{table}
\begin{table}
\caption{Properties of GCs in KKs\,3 and ESO\,269-66.} 
\label{tab:prop_gcs}
\begin{center}
\begin{tabular}{lrr}
\hline\hline
               & GC in KKs\,3 & GC in ESO\,269-66           \\
\hline
$\rm (V-I)_0$, mag  & 0.90$\pm0.06^a$ & 0.93$^d$  \\
$\rm M_V$, mag      & -8.48$^b$& -9.9$^d$   \\
$\rm R_h$,pc        & $4.8\pm0.2^a$  & $2.5\pm0.13^d$    \\
$\feh$,dex      & -1.55$\pm0.2^{a}$ & -1.5$\pm0.2^{a}$  \\
Age,Gyr         & 12.6$\pm1.5^{a}$ & 12.6$\pm1.5^{a}$ \\
$\rm V_h$, km~s$^{-1}$     & 316$\pm$7$^c$ & 774$\pm$6$^a$ \\
\hline
\end{tabular}
\end{center}
\medskip
$^a$: This work, $^b$: Karachentsev et al. (2015b), $^c$:~Karachentsev et al. (2015a) , 
$^d$: Georgiev et al. (2009).
\end{table}
KKs\,3 is a unique highly isolated dSph recently discovered by Karachentsev et al. (2015b).
It is one of the faintest known field galaxies (Karachentsev et al. 2015a). 
The arguments are:
the colour-magnitude diagram (CMD) of the object consisting only of old (age$\rm >1$~Gyr) stars, 
the absolute magnitude $\rm M_V\!=-12.6$~mag, effective surface brightness 
$\rm SB_{Ve}\!=\!24.9$~mag/arcsec$^{-2}$, 
neutral hydrogen mass-to-stellar mass ratio ($\rm M_{HI}/M^*\!=\!10^{-2.34}$)
estimated by the luminosity of the galaxy in the K-band
and star formation rate per unit galactic luminosity (specific star formation rate)
$\rm sSFR\!=\!10^{-13.6}$ 
estimated from the ultraviolet flux (Karachentsev et al. 2015a). 
For comparison, $sSFR$ of a dwarf irregular galaxy 
having similar luminosity in the B-band is 2-2.5 orders of magnitude larger.
The mass of the galaxy is $\rm 2\times10^7 M_{\sun}$ (Karachentsev et al. 2015b). 
The nearest massive neighbour to KKs\,3 is M31 at a distance of $\sim 1.7$~Mpc.
A long-slit absorption-line spectrum of the GC in KKs\,3 which was taken with the RSS at the SALT in 
bright time in January 2015 allowed us to
estimate radial velocity of the galaxy $\rm V_h\!=\!316 \pm 7$~km~s$^{-1}$ (Karachentsev et al. 2015a).
There was no neutral hydrogen detected in KKs\,3. The $\rm sSFR$ and $\rm M_{HI}/M^*$ parameters
presented by Karachentsev et al. 2015a indicate that the galaxy has transformed most 
of its gas into stars.

Star formation histories of ESO\,269-66 and KKs\,3 look similar.
Three star forming bursts likely occurred in KKs\,3 12--14 , 4--6,
and 0.8--2 Gyrs ago (Karachentsev et al. 2015b), where
the oldest one was the most powerful with mean $\feh=-1.9$~dex.
 Most stars (74\%) of the galaxy were formed {\akn during} this period (Karachentsev et al. 2015b).

We present the description of spectroscopic observations for the
GCs in ESO\,269-66 and KKs\,3 in Section~\ref{observations}.
In Sections~\ref{method} and \ref{results}, we describe our methods developed for
determination of ages, helium abundance (Y), \feh\, and the abundances of different chemical elements
using medium-resolution optical spectra and present our results. 
We compare the spectra of the studied two nuclei with medium-resolution spectra of Galactic GCs 
from the library of Schiavon et al. (2005, hereafter Sch05)\footnote{
The library by Sch05 contains high signal-to-noise
integrated-light spectra of 40 Galactic GCs with $FWHM\sim3.1$\AA\ resolution including
spectra of nine GCs with $\feh\sim-1.6$~dex.} 
and find possible Galactic analogues for our sample nuclei of dSphs
(Sections~\ref{compar_abund}, \ref{compar_age}). 
 We analyse CMDs of the selected Galactic GCs in Section~\ref{compar_CMDs}.
Structural and photometric parameters for the GC in KKs\,3 are presented in Section~\ref{structure}.
Section~\ref{Discussion} addresses the question: 
whether the GCs in KKs\,3 and ESO\,269-66 host multiple stellar populations
and what the origin of KKs\,3 and its nucleus is.

\section{Observations and data reduction}
\label{observations}
The spectroscopic observations were carried out with
SALT (Buckley et al. 2006, O\arcmin~Donoghue et al. 2006) in the period of 
15 January 2015 - 9 July 2015 (Table~\ref{tab:obslog})
with the Robert Stobie Spectrograph (RSS; Burgh et al. 2003, Kobulnicky et al. 2003).
We used the long-slit mode with a slit width 1.25\arcsec, the grating pg0900,
blocking filter pc03400 and the Camera Station parameter 26.5. The spectra cover a
spectral range of 3700--6700~\AA\ with a reciprocal dispersion of 0.97 \AA\ pixel$^{-1}$
and spectral resolution of $\rm FWHM=5$\AA.
A 30-minute exposure in the same instrumental mode of the GC in ESO\,269-66 was made during
the commissioning time in July 2011. 
Primary reduction of the data was done with 
the SALT science pipeline (Crawford et al. 2010). The reduction of the SALT long-slit data 
was done in the way described by Kniazev et al. (2008).
\begin{table}
\caption{Journal of  spectroscopic observations.} 
\label{tab:obslog}
\begin{center}
\begin{tabular}{lrrr}
\hline\hline
Object &  Date &  $T_{exp}$ & Seeing  \\
       &       &   (s)      & (arcsec) \\
\hline
KKs\,3   & 09/07/2015   & 3$\times$900  & 1.6-1.8 \\
       & 15-17/01/2015& 5$\times$1000 &  1-2    \\
E269-66& 11/05/2015   & 3$\times$900  &  $\sim$1     \\
       &  1/06/2011   & 2$\times$900  &  $\sim$1     \\
\hline
\end{tabular}
\end{center}
\end{table}
To improve the signal-to-noise ratio, the resulting one-dimensional spectra
were smoothed using the adjacent-averaging method with five points window.
The final signal-to-noise ratios in the resulting smoothed summary spectra of GCs in KKs\,3 and ESO\,269-66
 reach $\sim$100 at 5000 \AA.

\section{Analysis of the spectra}
\label{sec_analysis}
\subsection{Method}
\label{method}

\begin{figure}
\centering
\includegraphics[angle=-90,scale=0.32]{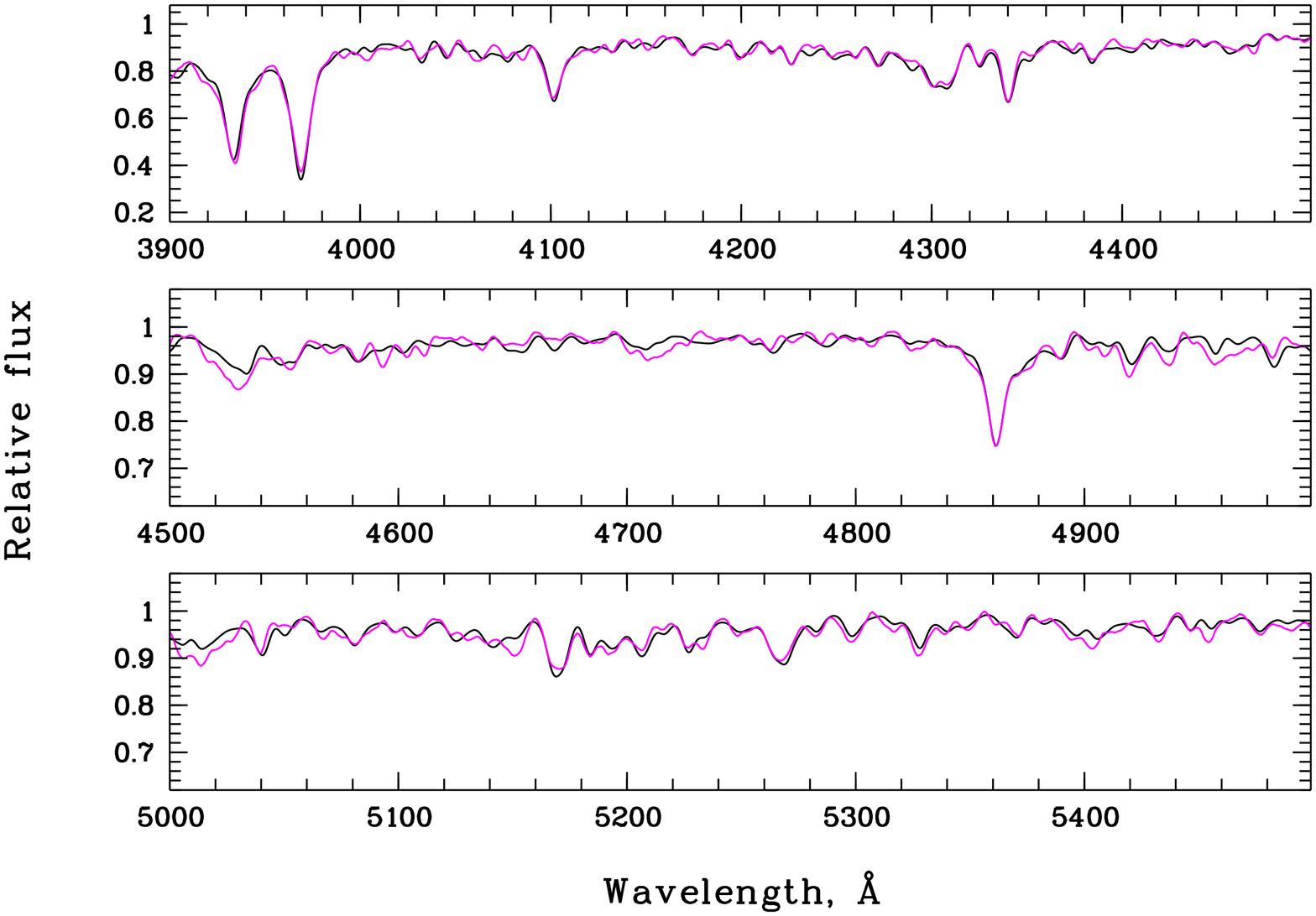}
\includegraphics[angle=-90,scale=0.32]{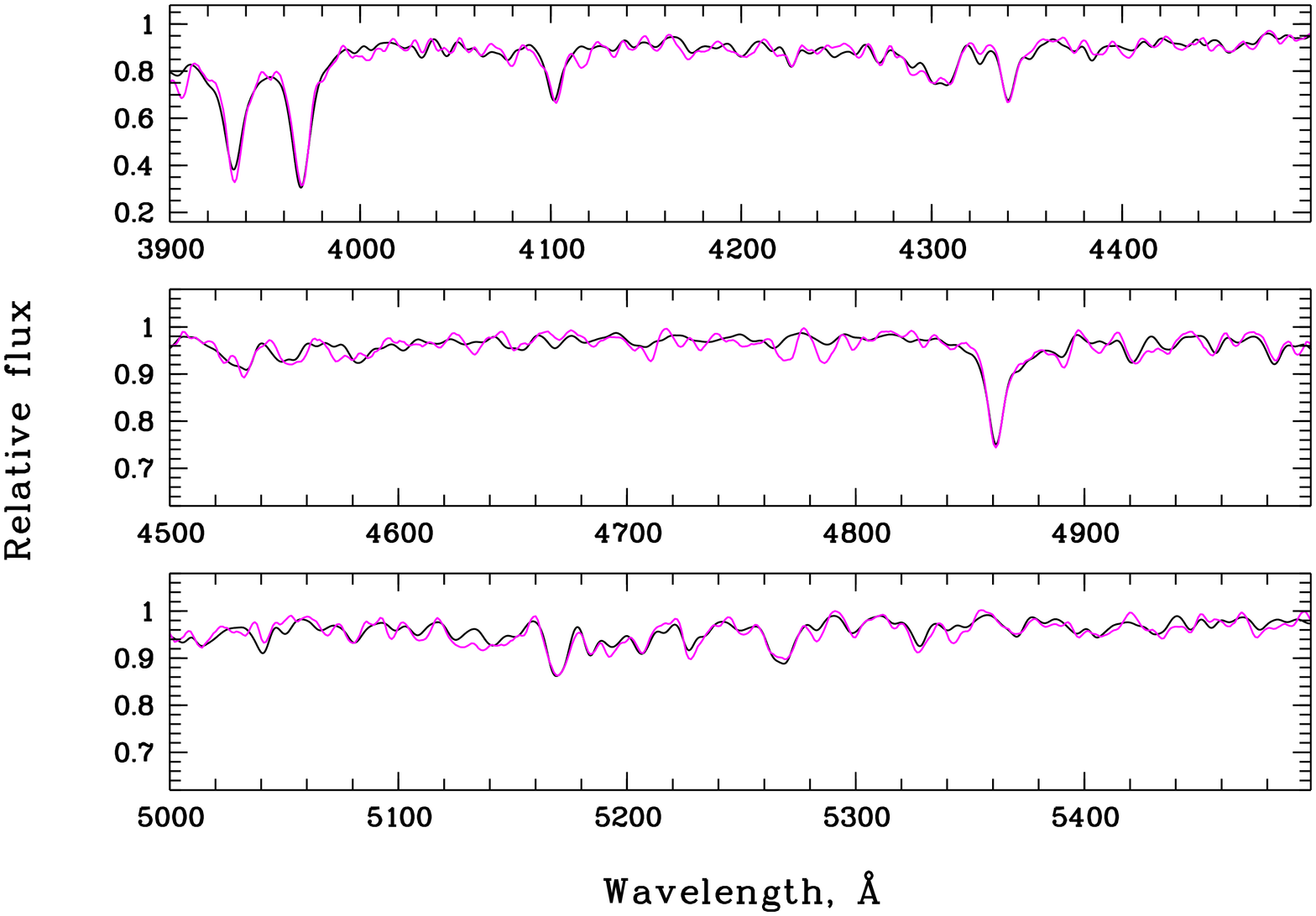}
\caption{Continuum normalized rest-frame spectra of GCs in ESO\,269-66 (top) and KKs\,3 (bottom) 
in comparison to the model ones (black lines) 
calculated according to the elemental abundances described in Section~\ref{sec_analysis} and Table~\ref{tab:abund}.}
\label{fig_KKs3_E269model}%
\end{figure}

The final medium-resolution spectra of the GCs in two
dSphs are shown in Fig.~\ref{fig_KKs3_E269model} in comparison to
the best-fitting model spectra computed as it is described below.

We used a program named {\it CLUSTER} (Sharina et al. 2013, 2014; Khamidullina et al. 2014) 
to calculate the synthetic integrated-light spectra for the GCs in ESO\,269-66 and KKs\,3.
Appendix~A in the paper by Sharina et al. (2014) explains the procedure and main principles
of synthetic stellar spectra computation using models of stellar atmospheres.
Synthetic spectra of stars were calculated according to 
their parameters: [Fe/H], effective temperature, and surface gravity. 
These parameters were set by the scaled solar stellar evolutionary isochrones by Bertelli et al. 
(2008, hereafter B08), which include the horizontal branch (HB) and AGB 
stages and various helium content. 
These ingredients are necessary, in particular, to explain HB morphologies and splitting of evolutionary sequences of GCs 
in several distinct branches (B08 and references therein). 
We did not interpolate theoretical isochrones but used the available set.
The calculated synthetic spectra of individual stars were summed according to the mass function by Chabrier (2005).
The computation of the synthetic spectra takes into account about 600000 atomic and 1800000 $^{12}$CH, $^{13}$CH,
and SiH molecular lines from the lists of Kurucz (1994) and Castelli \& Kurucz (2003).
The data for ten following molecules were kindly provided to us by Ya.V. Pavlenko:
CN, VO, TiO, SO, SiO, NO, MgO, CO, and AlO. Calculations of the abundances of all elements except 
for H, Ca, and Mg were performed under the assumption of local thermodynamic equilibrium (LTE).
Non-LTE effects were taken into account for  H, Ca, and Mg as it was described by Sharina et al. (2013).
The resulting model spectra of the GCs were normalized by the theoretical continuum and
broadened to the resolution of the observed spectra. 
We determined chemical abundances by fitting the medium-resolution observed spectra 
with the computed models.
To approximate the shape of the continuum in the observed spectra, we 
first smoothed the spectra by replacing each pixel value by 
the maximum of all points in the window $\rm 2\cdot R+1$, where $\rm R = 10\cdot FWHM$.
Then we implemented the running mean 
with the radius of $\rm 10\cdot FWHM$ to the filtered spectrum.

Ages and mean helium abundance (Y) for the GCs were determined by comparison of the observed
and the model Balmer line profiles. The influence of these two parameters
is not equivalent, which makes it possible to derive them simultaneously.
The temperatures of the Main Sequence turn-off (MSTO) stars become higher with the decreasing age.
This means that the depths of the cores and wings of the Balmer lines simultaneously strengthen. 
Increasing the Y values results in higher luminosities of hot HB stars. 
At the same time, temperatures and luminosities of stars populating Main Sequence (MS), 
sub-giant and red-giant branches do not change noticeably with the increasing of Y only and 
leaving all the other parameters unchanged (B08).
Pressure broadening is not that significant in the atmospheres of HB stars,
therefore, increasing of mainly Y results in strengthening of the central parts of Balmer lines
in the integrated spectra of GCs. The depths of H$\delta$, H$\gamma$, and H$\beta$
change differently with the change of Y, because hot HB stars contribute mainly to the blue part of the spectrum.
The method is illustrated in Figs~\ref{fig_ageY}, \ref{fig_ageYadd} and \ref{fig_ageYGCs} in
Appendix~\ref{app_age_he}
(please, see also Sharina et al. 2013, 2014; Khamidullina et al. 2014).
\begin{figure*}                                                                                                                                          
\hspace{-1cm}                                                                                                                                            
\centering                                                                                                                                               
\includegraphics[angle=-90,scale=0.61]{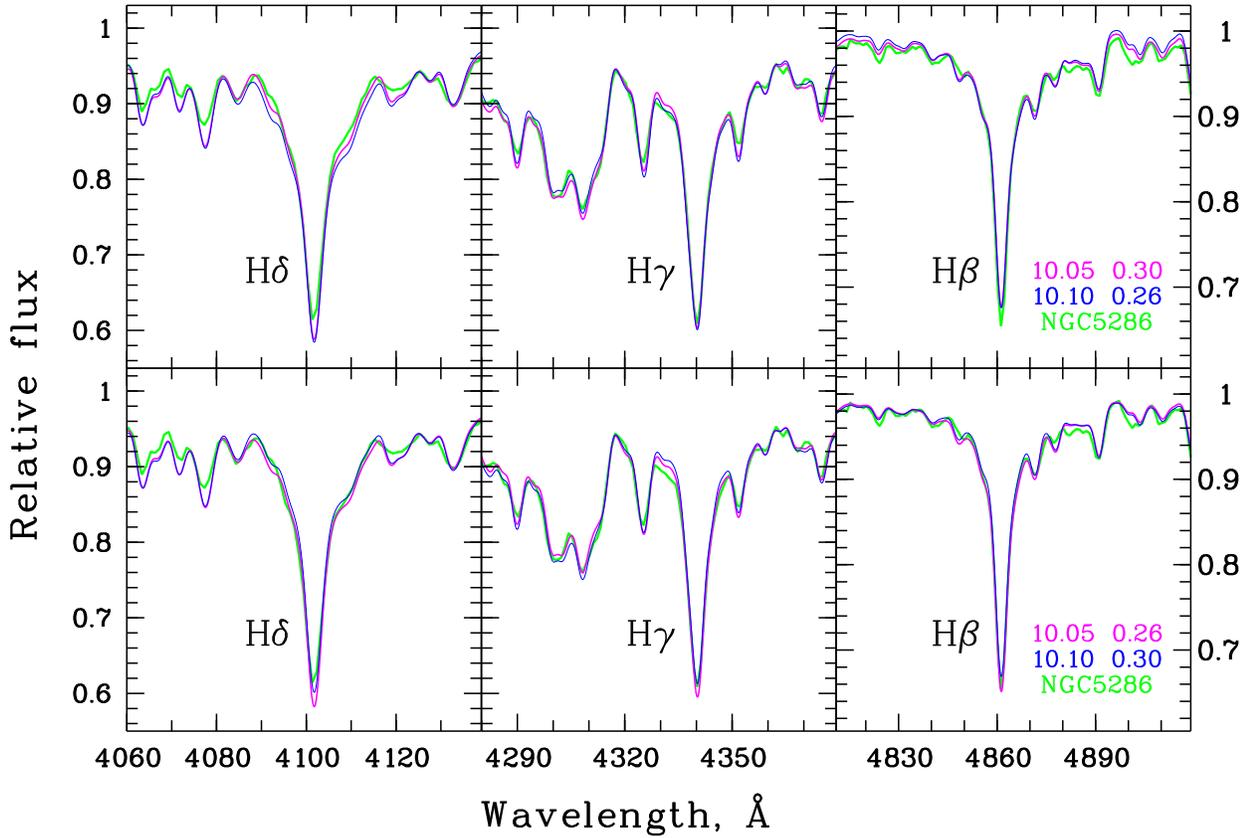}                                                                                                      
\hspace{-0.5cm}                                                                                                                                          
\caption{Comparison of three Balmer lines in the spectrum of NGC\,5286 and computed synthetic integrated-light spectra.                                   
We used for the calculation the same elemental abundances valid for NGC\,5286                                                                             
(Table~\ref{tab:abund}) and the isochrones by Bertelli et al. (2008) of the same metallicity $\rm Z=0.0004$ and                                           
different ages and specific helium abundances: 
$\rm log(Age)=10.05, Y=0.30$; $\rm log(Age)=10.10, Y=0.26$; $\rm log(Age)=10.05, Y=0.26$, and $\rm log(Age)=10.10, Y=0.30$.
It can be seen that the last case provides a better agreement with the spectrum of NGC\,5286}
\label{fig_ageY}%
\end{figure*}
\begin{figure}
\centering
\includegraphics[angle=-90,scale=0.32]{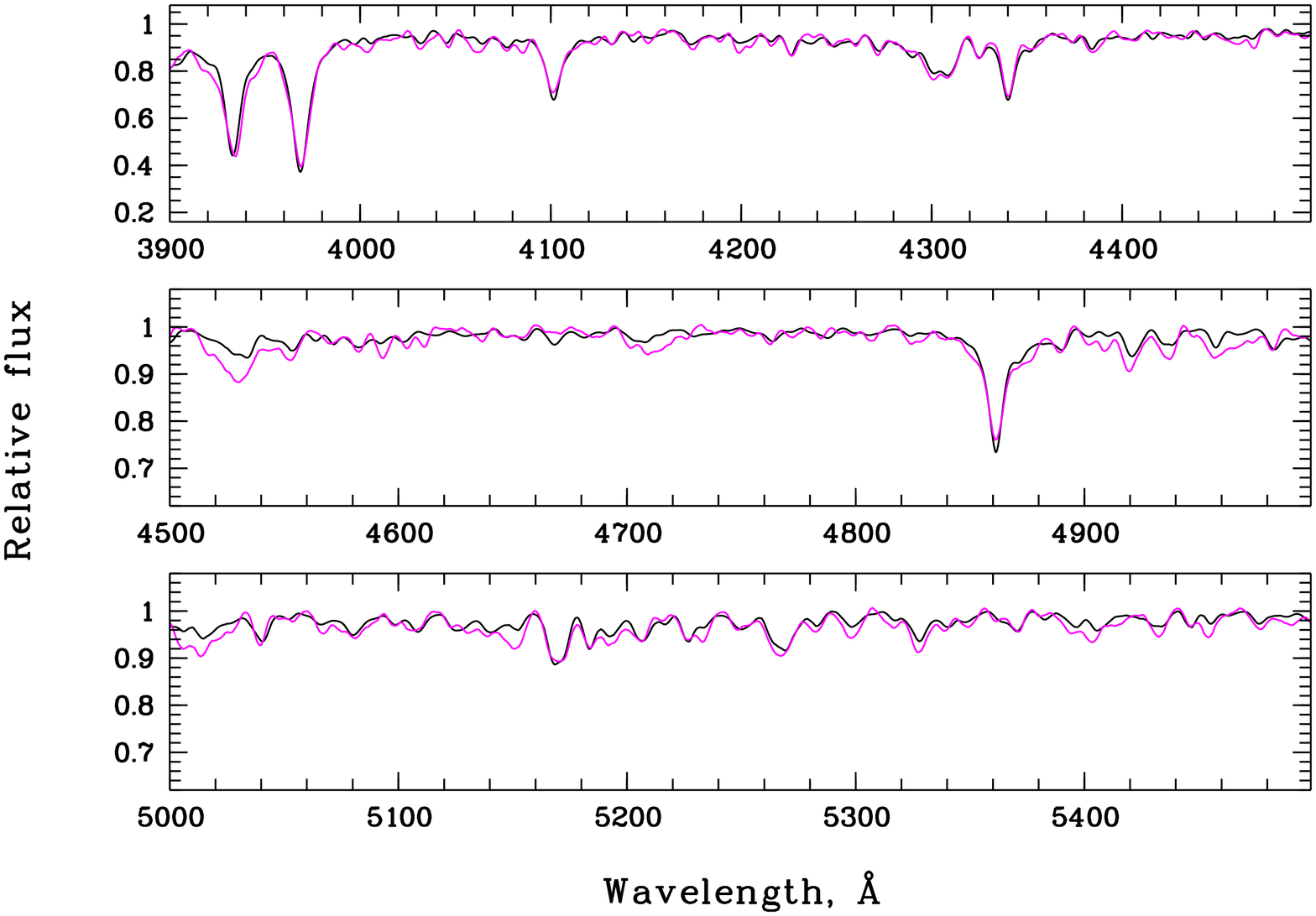}
\includegraphics[angle=-90,scale=0.32]{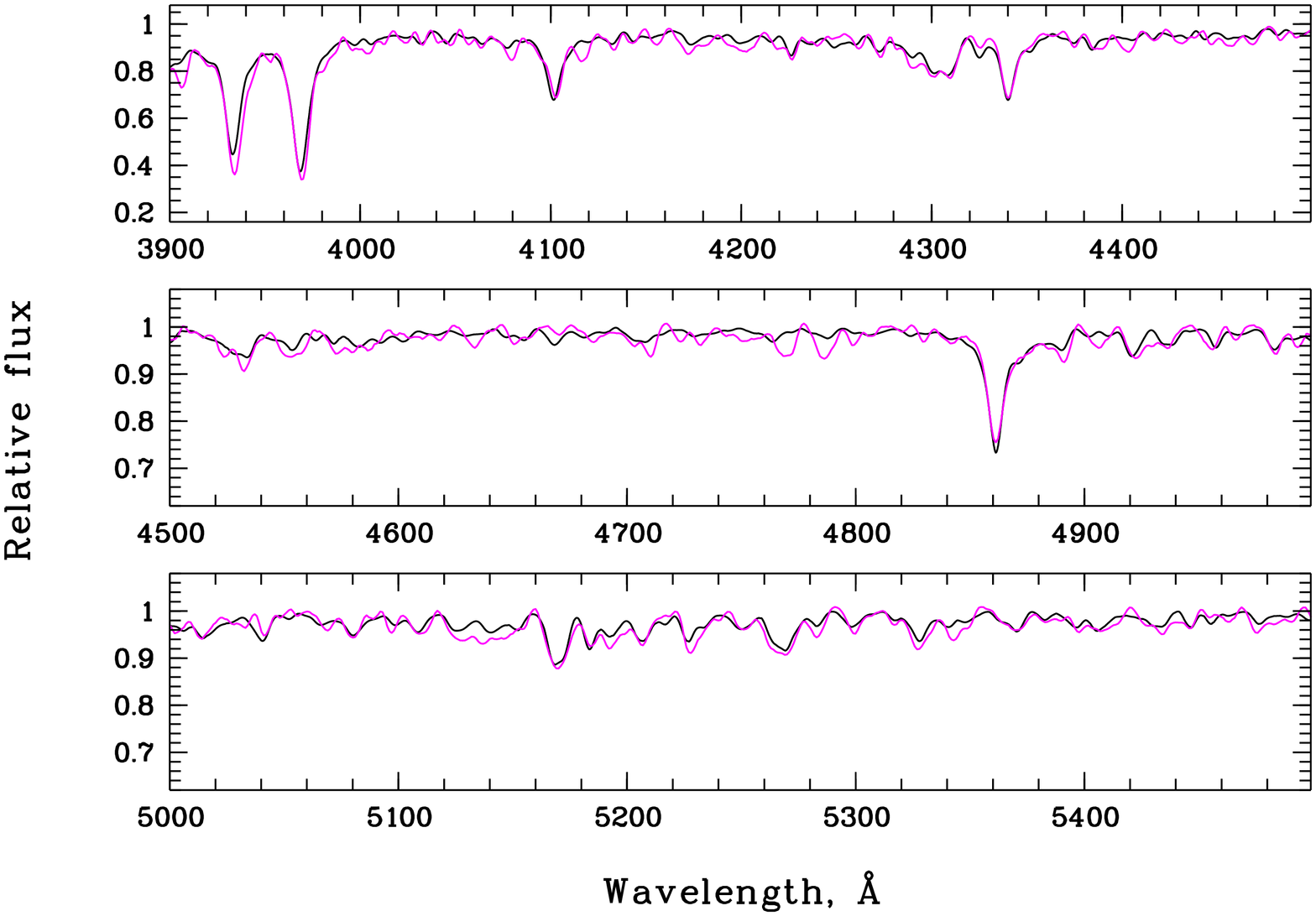}
\caption{Continuum normalised spectra of GC in ESO\,269-66 (top) and KKs\,3 (bottom) in comparison to the spectrum of 
NGC\,5286 (black line) from Schiavon et al. (2005).}
\label{figE269_KKs3_N5286}%
\end{figure}

Almost all spectroscopic lines are blended at a resolution of FWHM$\sim5$~\AA.
To derive chemical abundances we selected dominant
features in the spectra mostly sensitive to the abundances of the specific spectroscopic elements.
Iron gives a great number of spectral lines which allows its abundance to be derived with high accuracy.
The depths of iron lines in the integrated synthetic and observed spectra of the GCs depend on \feh\ and
micro-turbulent velocity ($\xi_{turb}$) (Figs~\ref{fig_XiTurbFe}, \ref{fig_XiTurbMg} in Appendix~\ref{app_elements}). 
The last parameter in its turn 
depends on metallicities, $\rm log~g$, and $\rm T_{eff}$ of stars within the cluster 
(Malavolta et al. 2014 and references therein).
The intensities of different iron lines are influenced by $\xi_{turb}$ in a different rate 
(Figs~\ref{fig_XiTurbFe}, \ref{fig_XiTurbMg} in Appendix~\ref{app_elements}). 
This fact allows us to select an optimum constant value of micro-turbulent velocity
for calculation of synthetic spectra of GCs using stellar atmosphere models.
The range of possible $\xi_{turb}$ values for low-metallicity stars is $1.6 \div 2.2$~km~s$^{-1}$ 
(e.g. Malavolta et al. 2014).
To constrain \feh\ and $\xi_{turb}$ at the resolution FWHM$\sim5$~\AA\ one can use, for example,  
such dominant blends as $\lambda\sim$5456, 5446, 5405, 5327, 5283, 5269, 5227, and 5216~\AA\ 
with the shapes mostly changed
by varying Fe and $\xi_{turb}$ (please, see also a list of the most important atomic lines
at the medium resolution in Fig.~2 by Dias et al. 2015).
Given the similarity of the analysed spectra, we choose a constant $\xi_{turb}$ 
value of $1.8$~km~s$^{-1}$ for all our sample GCs.
 When we used the relation from Marino et al. (2008),
$\rm \xi_{turb} = -0.254\cdot log~\!g +1.93$ (Malavolta et al. 2014), we obtained
the following corrections to the elemental abundances in dex: $\Delta \feh=+0.08$, $\Delta \cfe=-0.03$, 
$\Delta [Mg/Fe] = -0.04$, $\Delta \cafe = -0.04$, $\Delta \tife = -0.04$, $\Delta \crfe = -0.01$.
It can be seen that these corrections are less than the errors of our analysis.
\begin{table*}
\caption{Elemental abundances and their uncertainties for the GCs in KKs\,3 and E269-66
and for five Galactic GCs according to our measurements using medium-resolution integrated-light spectra (Section~\ref{sec_analysis}).}
\label{tab:abund}
\begin{center}
\begin{tabular}{lrrrrrrr}
\hline\hline 
GC                    &\feh          &  \cfe            & \nfe         &\mbox{[Mg/Fe]}& \cafe         & \tife        &  \crfe          \\ 
                      & (dex)        & (dex)            & (dex)        &(dex)         & (dex)         & (dex)        & (dex)           \\ 
\hline                                                                                                                                                
in E269-66            & -1.50$\pm0.2$& -0.12$\pm$0.2    & 0.25$\pm$0.4 &0.10$\pm$0.2  & 0.00$\pm$0.2  & 0.15$\pm$0.3 &  0.15$\pm$0.3   \\ 
  \noalign{\smallskip}                                                                                                  %
in KKs\,3               & -1.55$\pm0.2$& -0.18$\pm$0.2    & 0.45$\pm$0.4 &0.15$\pm$0.2  & 0.30$\pm$0.2  & 0.25$\pm$0.3 & 0.05$\pm$0.3    \\ 
\noalign{\smallskip}                                                                                                    %
 NGC\,1904             & -1.75$\pm0.2$& 0.03$\pm$0.1     & 0.30$\pm$0.3 &0.25$\pm$0.1  & 0.15$\pm$0.1  & 0.15$\pm$0.2  & 0.25$\pm$0.2   \\ 
 NGC\,5286             & -1.75$\pm0.2$& 0.02$\pm$0.1      &0.35$\pm$0.3 & 0.38$\pm$0.1 & 0.25$\pm$0.1  & 0.15$\pm$0.2  &  0.25$\pm$0.2  \\ 
 NGC\,6254             & -1.65$\pm0.2$& -0.15$\pm$0.1    & 0.25$\pm$0.3 &0.14$\pm$0.1  & 0.12$\pm$0.1  & 0.45$\pm$0.2  & -0.09$\pm$0.2  \\ 
 NGC\,6752             & -1.75$\pm0.2$& -0.06$\pm$0.1    & 0.30$\pm$0.3 &0.22$\pm$0.1  & 0.25$\pm$0.1  & 0.20$\pm$0.2  & 0.20$\pm$0.2   \\ 
 NGC~7089             & -1.75$\pm0.2$& -0.01$\pm$0.1    & 0.30$\pm$0.3 &0.35$\pm$0.1  & 0.15$\pm$0.1  & 0.20$\pm$0.2  & 0.20$\pm$0.2   \\ 
\hline
\end{tabular}
\end{center}
\end{table*}

After fitting \feh\ and $\xi_{turb}$, we focused at the abundances of chemical 
elements which have prominent lines and blends in the spectroscopic range.
Their depths may change by more than 1--3\% of the continuum level
while varying the corresponding abundances by 0.1--0.3~dex. 
These elements (and the corresponding spectroscopic features) are:
calcium (CaII H$+$K; CaI $\lambda=4226$~\AA), magnesium (MgH molecule $\lambda \sim4980\div5180$~\AA, 
Mg~I line 5183~\AA), chromium (Cr lines $\lambda \sim5207$, 5298~\AA), 
nitrogen plus carbon (CN molecule $\lambda \sim4120\div4220$~\AA, carbon (CH 4300~\AA\ band)
(Figs~\ref{fig_Ca} -- \ref{fig_Cr} in Appendix~\ref{app_elements}).
The depth of the last spectroscopic feature decreases with
an increase of oxygen abundance, but the amount of this change is $\sim$5 times less than 
the line depth variation for carbon (e.g. Lardo et al. 2016). 
The abundance of oxygen was set $\ofe \sim 0.5$~dex.
According to our calculations, this value provides the correct shapes and depths of the CH-band 4300~\AA\
and CN-band at $\lambda \sim4120\div4220$~\AA                                                                                                                                              

\subsection{Results}
\label{results}
\begin{table*}
\caption{Elemental abundances in dex from literature high-resolution spectroscopic studies of red 
giant branch stars in five Galactic GCs analogous to the GCs in KKs\,3 and ESO\,269-66.}
\label{tab:abundhr}
\scriptsize
\begin{center}
\begin{tabular}{llllllllllll}
\hline\hline 
NGC                  & \feh &  \cfe          & \nfe           & \ofe         & \nafe         &  \mbox{[Mg/Fe]}& \sife         & \cafe         & \tife         &  \crfe                \\ 
\hline                                                                                                                                                       %
\noalign{\smallskip}                                                                                                                                         %
1904$^b$             & -1.58$\pm$0.12 &  --            &  --           & 0.10$\pm$0.19& 0.32$\pm$0.25 & 0.26$\pm$0.07& 0.28$\pm$0.03 & 0.22$\pm$0.04 & 0.22$\pm$0.10 & -0.28$\pm$0.14           \\ 
\noalign{\smallskip}                                                                                                                                         %
5286$^a$             & -1.70$\pm0.07^b$ &  --            & --            & 0.44$\pm$0.22& 0.34$\pm$0.21 & 0.55$\pm$0.11&  0.40$\pm$0.09&  0.31$\pm$0.06& 0.33$\pm$0.10&-0.05$\pm$0.20            \\ 
\noalign{\smallskip}                                                                                                                                         %
6254$^b$             & -1.53$\pm$0.06 & -0.77$\pm$0.37 & 1.01$\pm$0.45 & 0.23$\pm$0.24& 0.17$\pm$0.27 & 0.44$\pm$0.13& 0.28$\pm$0.07 & 0.33$\pm$0.11 & 0.26$\pm$0.12 &~0.01$\pm$0.15           \\ 
\noalign{\smallskip}                                                                                                                                         %
6752$^b$             & -1.53$\pm$0.16 & -0.45$\pm$0.37 & 0.93$\pm$0.63 & 0.26$\pm$0.25& 0.32$\pm$0.26 & 0.38$\pm$0.15& 0.47$\pm$0.19 & 0.31$\pm$0.09 & 0.20$\pm$0.11 & -0.13$\pm$0.12            \\ 
\noalign{\smallskip}                                                                                                                                         %
7089$^c$             & -1.64$\pm0.08^b$ & -0.62$\pm0.14^b$& --           & 0.42$\pm0.16$& 0.06$\pm0.23$ & 0.38$\pm0.08$& 0.40$\pm0.01$ & 0.28$\pm0.02$ & 0.17$\pm0.02$ & -0.06$\pm0.03$ \\ 
\noalign{\smallskip}
\hline \hline
\end{tabular}
\end{center}
$^a$: Marino et al. (2015), $^b$: Roediger et al. (2014),                                                                                  
$^c$: Six canonical RGB stars (r-only) by Yong et al. (2014).                                                                              
\end{table*}                                                                                                                               
\begin{table*}
\caption{Elemental abundances in dex and their errors from literature high-resolution integrated-light 
spectroscopic studies for the two objects 
in common between our study and those of Colucci et al. (2017) (C17) and Larsen et al. (2017) (L17). 
Only \feh\ and the chemical elements common with the listed ones in Table~\ref{tab:abundhr} are shown.
In the third row of the table, we present FeI and TiI abundances from the paper by C17 (marked by asterisk).} 
\label{tab:abundil}
\scriptsize
\begin{center}
\begin{tabular}{lrrrrrrr}
\hline\hline 
NGC                 & \feh\mbox{, $\sigma$} & \nafe\mbox{, $\sigma$} &  \mbox{[Mg/Fe], $\sigma$}& \sife\mbox{, $\sigma$} &  \cafe\mbox{, $\sigma$}& \tife\mbox{, $\sigma$}&  \crfe\mbox{, $\sigma$}  \\ 
\hline                                                                                                           %
6254$^{L17}$            & -1.481, 0.034     & -0.038, 0.003          & 0.378, 0.067               &   --                & 0.380, 0.137            & 0.412, 0.097  &  0.104, 0.411 \\
\noalign{\smallskip}                                                                                       
6752$^{L17}$            & -1.883, 0.039     & 0.302, 0.023           & 0.391, 0.130               &   --                & 0.355, 0.114            & 0.342, 0.128  &  -0.101, 0.070  \\
\noalign{\smallskip}
6752$^{C17}$            & -1.58, 0.20$^*$   & 0.04, 0.07             & 0.06, 0.07                 &  0.40, 0.07         & 0.41, 0.04              & 0.39, 0.06$^*$ &  -0.09, 0.09 \\
\noalign{\smallskip}
\hline \hline
\end{tabular}
\end{center}
\end{table*} 
\begin{figure}
\centering
\includegraphics[angle=-90,scale=0.32]{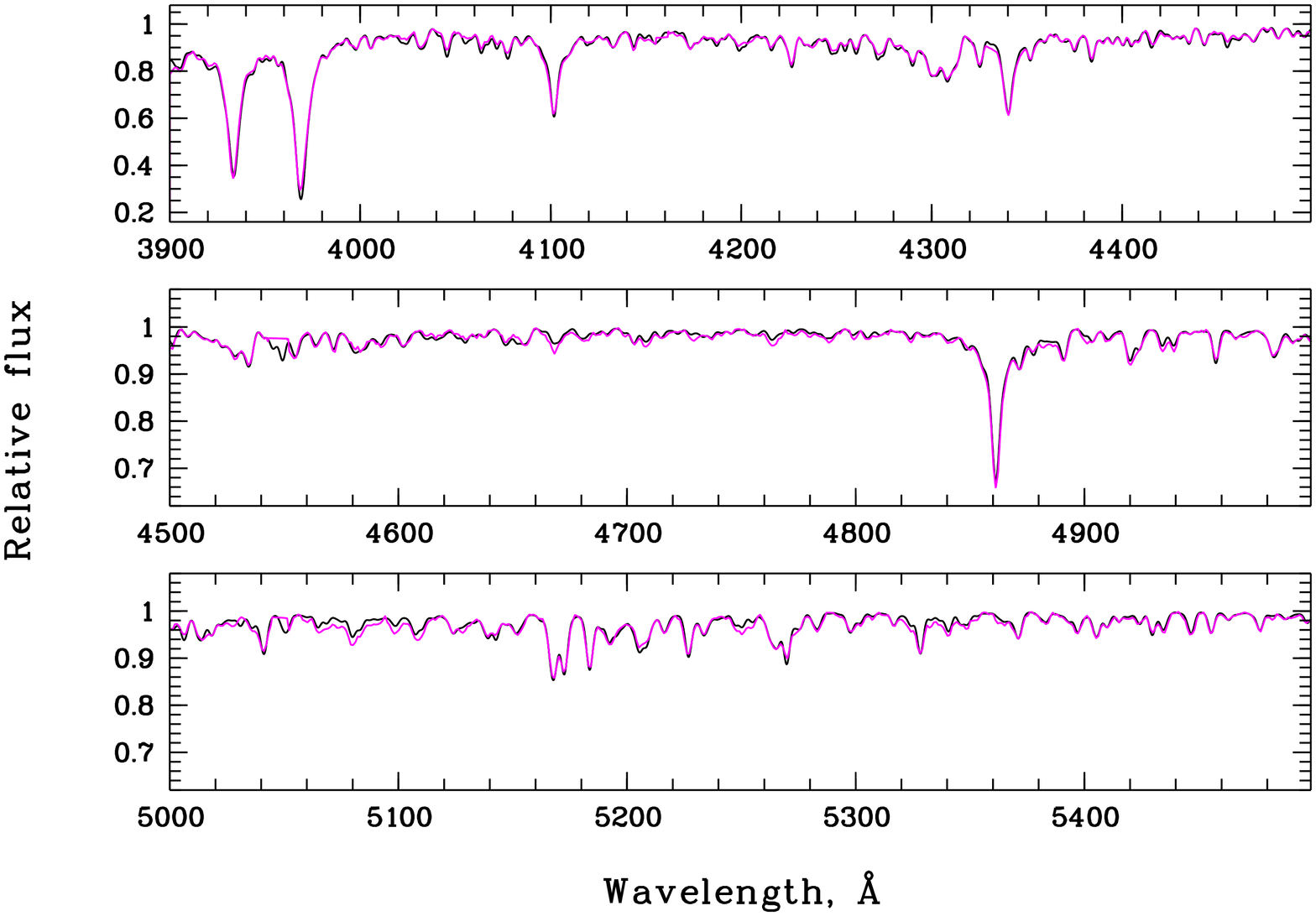}
\caption{Continuum normalised spectrum of NGC\,5286 from Schiavon et al. (2005) in comparison with the model one (black line)
(please, see Table~\ref{tab:abund} and Section~\ref{sec_analysis} for details).}
\label{fig_N5286model}%
\end{figure}

The derived chemical abundances for two extragalactic and a comparative sample of five 
Galactic GCs are presented in Table~\ref{tab:abund}.
We used the spectra of Galactic GCs presented by Sch05.
The isochrone providing the best fit of Balmer lines in the spectra of all the sample GCs
has the following parameters (B08): 
$\rm Z=0.0004$, $\rm Y=0.30$, and $\rm log(Age)=10.10$.
 The abundance analysis of the Galactic GCs was performed using the same procedure as
for two extragalactic GCs (Section~\ref{method}). 
Errors of the elemental abundances (Table~\ref{tab:abund})
include fitting uncertainties in an rms sense, signal-to noise effects,
and the errors occurring due to
the combined influence of several parameters on the spectrum
such as micro-turbulent velocity, abundances of Fe, and several other elements.
For instance, the strength of the CH-band at 4300~\AA\ depends mainly on the
 C and O abundances. To a lesser extent, it depends also on $\xi_{turb}$ and the 
 abundances of Fe, Ti, Ca, Mg, Al, and Si.

We compared spectra of two nuclei not only with the computed theoretical ones
but also with the medium-resolution spectra of Galactic GCs. 
It appears that spectra of five Galactic GCs from the library of Sch05 
 are very similar with the spectra of our two GCs in dSphs.
These objects are
NGC\,5286, NGC\,1904, NGC\,6254, NGC\,6752, and NGC~7089.
Fig.~\ref{figE269_KKs3_N5286} illustrates this comparison
using the spectrum of NGC\,5286 as an example. 
 Comparison of the spectrum of NGC\,5286 
with the corresponding best-fitting synthetic one 
computed using the elemental abundances listed in Table~\ref{tab:abund}
is shown in Fig.~\ref{fig_N5286model}.
Comparison of the NGC\,5286 spectrum to the spectra of NGC\,1904, NGC\,5986,
NGC\,6254, NGC\,6752, NGC~7089, and of several other
Galactic GCs from the library by Sch05 having $\feh\!\sim\!-1.6$~dex is 
presented in Fig.~\ref{fig_ageYGCs} (Appendix~\ref{app_age_he}) and 
at an anonymous ftp site\footnote{ftp://ftp.sao.ru/pub/sme/TwoDSphs/AppendixE.pdf}.

 Appendix~\ref{StelPhot} demonstrates HST stellar photometric data for the
GCs in ESO\,269-66 and KKs\,3 and for the surrounding stellar fields. 
Probable contamination of intermediate-age high-metallicity stars to the integrated light
of the GCs is considered. The presented analysis indicates that the contamination is small.
Unfortunately, the central regions of the GCs are not resolved into individual stars due to 
crowding effects which are stronger for the more distant ESO\,269-66.
The probability to find bright projected foreground stars is higher for the GC in ESO\,269-66,
because the surface density and total number of intermediate-age high-metallicity stars is higher
in this galaxy than that in KKs\,3.

Appendix~\ref{sec_lick} illustrates the similarity of absorption-line indices measured 
using flux-calibrated spectra in 
the Lick system (Burstein et al. 1984, Worthey et al. 1994, 
 Worthey \& Ottaviani 1997, Trager et al. 1998) for the sample extragalactic and Galactic GCs.
The indices for the GCs in KKs\,3 and E269-66 calculated using the calibration of 
 Katkov, Kniazev \& Sil'chenko (2015) are presented in Tables~\ref{lickind1} and \ref{lickind2}
  together with the corresponding measurement errors. The total index errors consist
of measurement errors and the errors of transformation to the Lick standard system.
 Lick indices for the Galactic GCs were taken from Schiavon et al. (2012).
Fig.~\ref{lick} shows two age-metallicity diagnostic plots of
hydrogen-line indices H$\delta$ and H$\beta$ versus the index 
$\rm \mgfe$~\footnote{$\rm \mgfe=\sqrt{Mgb (0.72 Fe5270 + 0.28 Fe5335)}$}
and four diagnostic plots with indices that are sensitive to iron, Mg, Ca, C, N, and O.
The $\alpha$-enhanced models by Thomas et al. (2003, 2004) are over-plotted.
Tables~\ref{lickind1} and \ref{lickind2} and Fig.~\ref{lick}
demonstrate that several Galactic GCs from  Sch05 
with metallicities $\feh\!\sim\!-1.6$~dex and the two GCs in dSphs show similar indices 
within the measurement errors.
      
\subsubsection{Comparison with literature data: abundances}
\label{compar_abund}
\begin{table}                                                                                                                             
\caption{Comparison between the derived in this study and literature alpha-element ratios and ages for Galactic GCs. 
The table shows age from VandenBerg et al. (2013, V13) and \afe\ from Carretta et al. 
(2010, C10 and references therein) where \afe\ 
is the average of $\rm [Mg/Fe]_{max}$, $\rm[Si/Fe]_{min}$, and \cafe.
Our age estimate is equal for all five GCs: $Age_{our}=12.6$~Gyr.} 
\label{tab:prop5GCs}
\begin{center}
\begin{tabular}{lccc}
\hline\hline                                                                                                                       
NGC    & $\afe^{C10}$  & {\small $(\cafe+\mgfe)^{our}/2$}& Age$^{V13}$      \\ 
       & {\small[dex]} & {\small[dex]}                   & [Gyr]            \\
\hline                                                                   
1904   & 0.31          & 0.20$\pm0.14$                   &  --            \\
5286   & --            & 0.32$\pm0.14$                   & 12.50$\pm0.38$ \\
6254   & 0.37          & 0.13$\pm0.14$                   & 11.75$\pm0.38$ \\
6752   & 0.43          & 0.24$\pm0.14$                   & 12.50$\pm0.25$ \\
7089   & 0.41          & 0.25$\pm0.14$                   & 11.75$\pm0.25$ \\
\hline
\end{tabular}
\end{center}
\end{table}
Elemental abundances from literature high-resolution spectroscopic studies for 
five Galactic GCs are listed in Tables~\ref{tab:abundhr} and \ref{tab:abundil}. 
It should be noted that 
our method yields systematically lower \feh\ values by 0.1--0.15 dex (Table~\ref{tab:abund}) 
for NGC\,1904, 5286, 6752, and 7089 than the corresponding metallicities obtained in high-resolution 
spectroscopic studies of red giant branch stars in these GCs (Table~\ref{tab:abundhr}) .
We suggest that this is the result of using
i) the same microturbulent velocities for all stars in a cluster and
ii) scaled-solar isochrones and model atmospheres.
The method may yield biased results in case of non-solar peculiar chemical patterns.
 NGC\,6254 shows a higher metallicity than other four Galactic GCs according to our analysis.
However, it is not clear whether this may be due to the influence of Galactic
field stars on the integrated spectrum of NGC\,6254 (Sch05). 
Foreground reddening of NGC\,6254 is strong. 
Additionally, a strong dust extinction gradient was detected within the cluster area (Leon et al. 2000).

Analysing the chemical compositions of the sample GCs, one has to take into account
that all five considered Galactic GCs host multiple stellar populations 
in the sense that their evolutionary sequences split in several components
characterized by different Na, C, O, and He abundances (Piotto et al. 2015, Marino et al. 2015, 
Milone et al. 2013, 2015),
and that their populations consist of several stellar generations characterized by
correlated variations of light elements (Marino et al. 2015; Carretta et al. 2009, 2010;
Fabbian et al. 2005). Variations of the abundances of Fe and s-process elements 
were detected in stars of NGC\,5286 and NGC~7089
(Milone et al. 2015, Marino et al. 2015). 

The following conclusions could be derived from the comparison of Tables~\ref{tab:abund}
and \ref{tab:abundhr}:
i) average abundances of RGB stars determined in high-resolution studies 
for a given element are similar within the errors for all five Galactic GCs;
ii) when taking into account the measurement errors and a relatively small number of the RGB stars 
analysed in high-resolution studies, {\it the mean} medium-resolution and high-resolution
abundances for a given element 
look similar for all five Galactic GCs for all of the considered elements except C and N; 
iii) abundances of Mg, Ca, Ti, and Cr look similar within the errors for the
Galactic and two extragalactic GCs.
Actually, there are slight systematic discrepancies between high and medium-resolution abundances of 
Mg and Ca for the Galactic GCs in Tables~\ref{tab:abund} and \ref{tab:abundhr},
although, the values of these differences are 
comparable with the corresponding errors of the abundance analysis.
On average, RGB stars in our sample GCs appear to be more abundant in the light elements
than the clusters' integrated light. A much stronger discrepancy was 
discovered for magnesium by Colucci et al. (2017).

The differences in the C and N abundances  between RGB stars (Table~\ref{tab:abundhr}) and
the whole cluster (Table~\ref{tab:abund}) may be interpreted as the result of an internal stellar evolution.
The phenomenon of mixing within the atmospheres
of RGB stars has long been studied (e.g. Kraft 1984, Gratton et al. 2012). 
The material processed through the CNO cycle in the stellar interiors brings up to the photosphere 
by convective processes such as, for example, thermohaline mixing (Charbonnel \& Zahn 2007). 
In this way, C and O are depleted and N is enhanced relative to stars
on the Main Sequence, or the sub-giant branch. 
Oxygen is depleted to a much lesser extent than carbon (Carretta et al. 2005).
According to the Chabrier (2005) mass function used by us in the calculation of the synthetic spectra of GCs, 
the radiation of RGB stars and that of MSTO stars comprise up to $\sim 30 \% $ each
of the total cluster optical light (please, see also Fig.~14 in Khamidullina et al. 2014).
Therefore, at first glance the mentioned discrepancies do not look surprising.
However, we measure C and N abundances average for a cluster using molecular bands
which are more intensive in RGB stars.
Please, note also the presence
of several generations of stars in the sample objects differing by C, N, O, Na, and He.
Pasquini et al. (2008) found an unpolluted MSTO star in NGC\,6752 having $\nfe=0.32\pm0.1$ and $\ofe=0.24$.
Unfortunately, there are no exhaustive data in the literature concerning the light-element abundances
for MS stars in these GCs.
The aforementioned slight differences between high and medium-resolution abundances of
Mg and Ca for the Galactic GCs in Tables~\ref{tab:abund}, \ref{tab:abundhr} 
may also be caused by mixing within the atmospheres of RGB stars.
Dissimilarities between abundances of chemical elements determined in high-resolution 
integrated-light analysis and spectroscopy of individual stars in GCs 
were revealed in the course of extensive studies of integrated-light spectra of many
 Galactic and extragalactic GCs (Colucci et al. 2017 and Larsen et al. 2017 and references therein).
Future high-resolution studies of integrated-light spectra and of many stars with 
different masses, $\rm T_{eff}$ and $\rm log(g)$ in GCs will finally establish the values
of the dissimilarities for different chemical elements.

\subsubsection{Comparison with literature data: ages and Y}
\label{compar_age}
\begin{table*}
\caption{Contribution of HB stars to the integrated light of the cluster in the B-band
with respect to the corresponding contribution of RGB stars brighter than the level of HB 
($\rm I_{HB}/I_{RGB}$).
These values are shown for five Galactic GCs (CMDs in Fig.~\ref{figproportion}) 
and for synthetic spectra of GCs computed using isochrones from B08:
(1) $\rm Z=0.0004, log(age)=10.1, Y=0.3$, 
(2) $\rm Z=0.0004, log(age)=10.05, Y=0.3$,
(3) $\rm Z=0.0004, log(age)=10.1, Y=0.26$ and (4) $\rm Z=0.0004, log(age)=10.05, Y=0.26$ 
(please, see text for details).
The stars belonging to HBs and RGBs of Galactic GCs were selected
as it is demonstrated in Fig.~\ref{figproportion}.} 
\label{tab:proportion}
\begin{center}
\begin{tabular}{lccccccccr}
\hline\hline
    &  NGC\,1904   &  NGC\,5286    & NGC\,6254   & NGC\,6752 & NGC\,7089        &  1        &  2        &  3        &  4\\
\hline
$\rm I_{HB}/I_{RGB}$ &  $0.85$   & $0.79$     &$0.79$    & $0.8$  & $0.78$        &$0.77$    &$0.47$    & $0.65$   & $0.54$ \\
\hline
\end{tabular}
\end{center}
\end{table*}
\begin{figure*}
\centering
\includegraphics[angle=-90,scale=0.68]{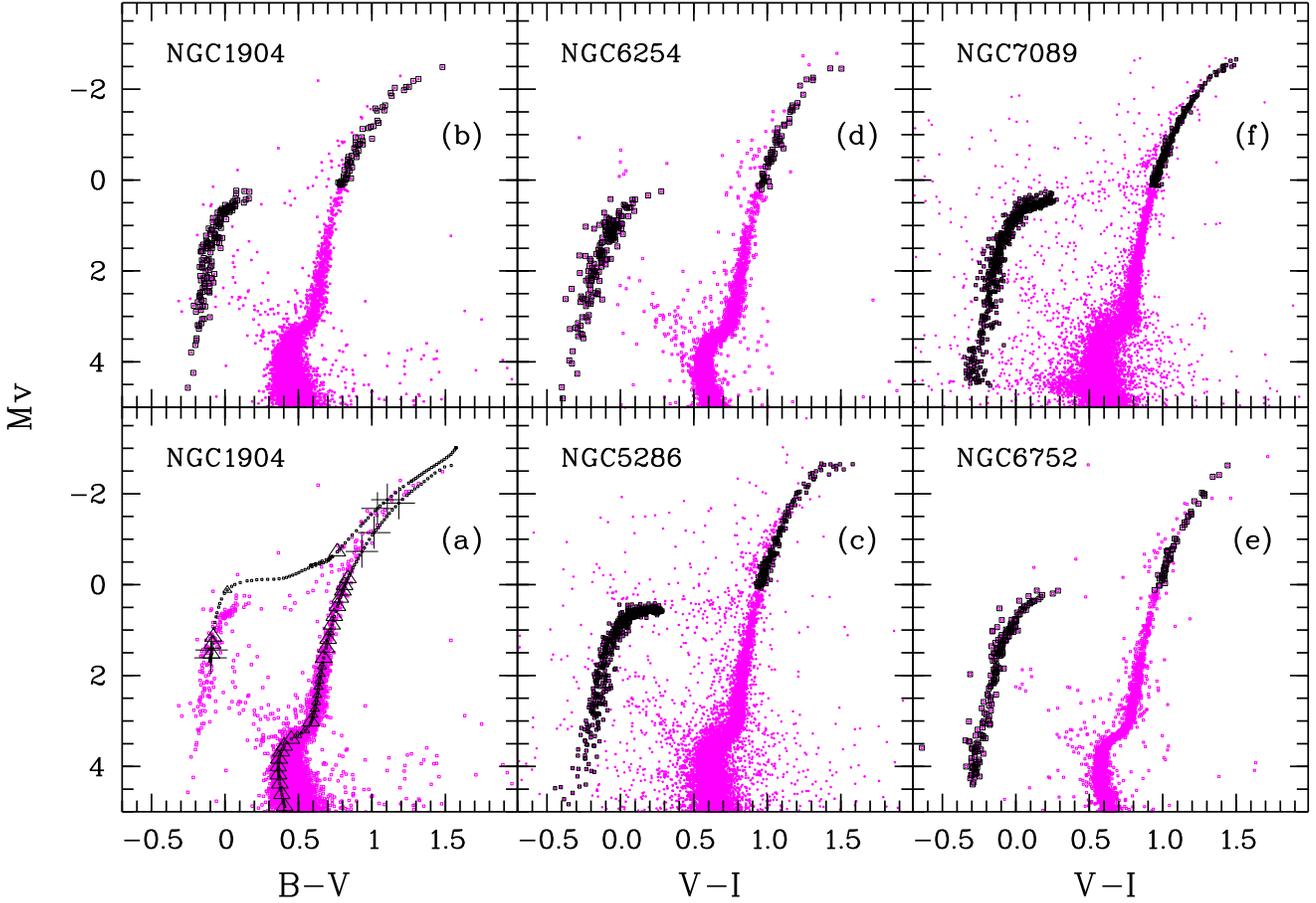} 
\caption{(a): CMD of NGC\,1904 (Piotto et al. 2002) and the overplotted isochrone 
$\rm Z=0.0004$, $\rm Y=0.30$ and $\rm log(Age)=10.10$ (B08).
Evolutionary stages contributing to the total cluster's light less than 1\%, 1--3\% and more than 3\%
are shown as small triangles, large triangles, and big crosses respectively 
(Section~\ref{compar_age}).
(b-e) Selection of HB and RGB stars using deep HST CMDs from Sarajedini et al. (2007)
and Piotto et al. (2002). }
\label{figproportion}%
\end{figure*}

The knowledge of absolute ages and helium content of GCs is important
for many reasons (e.g. VandenBerg et al. 2013, Charbonnel 2016).
The ages of old Galactic GCs are measured with an accuracy of 1--2 Gyr and are model-dependent.
Therefore, it would be instructive to compare the results of deep photometric and spectroscopic studies to
achieve better accuracies. Age and He abundance are the second and the third parameters (the first being 
metallicity) defining the distribution of stars on the HB of GCs (Gratton et al. 2010).
Helium content is one of the main factors influencing evolution of stars.
In particular, only about 30\% of stars in GCs at the ages of 12 Gyrs,
are the first generation stars.
Helium abundance is higher in stars belonging to the second and subsequent (if any) stellar generations in GCs. 
Large variations in He abundance in GCs are correlated with the variations of Na, O, 
and other elements (D'Antona et al. 2002). He-enriched stars evolve faster at a given age.
At the MSTO such stars are less massive than the He-poor ones.
``Variations in the median He abundance allow to explain the
extremely blue HB of GCs like NGC\,6254 ($=$M~10) and NGC\,1904 ($=$M~79)...''(Gratton et al. 2010).

The literature \afe\ values from Carretta et al. (2010) and ages from VandenBerg et al. (2013)
are summarised in Table~\ref{tab:prop5GCs} together with our results.
 Carretta et al. (2010) used the average of the maximum Mg abundance ($\rm[Mg/Fe]_{max}$), minimum Si
abundance ($\rm[Si/Fe]_{min}$), and \cafe\ as a measure
of the over-abundance of $\alpha$-elements, because this parameter coupled with \feh\ and several other abundances 
describes well the chemical composition of the primordial population in GCs.
 It is seen that all the GCs have EHBs and old age ($\sim12$~Gyr) (e.g. VandenBerg et al. 2013).
Light-element content of their RGB stars is enhanced, $\afe\sim0.4$ 
(Carretta et al. 2010 and references therein).
Table~\ref{tab:prop5GCs} illustrates a pretty good agreement between the literature high-resolution 
spectroscopic and deep photometric data
and our results that are based on the analysis of medium-resolution spectra.

In the following, we will focus on some 
details of our procedure of age and Y determination
to better understand the reasons of the existing discrepancies between the data derived
using different methods (Table~\ref{tab:prop5GCs}).
Figure~\ref{fig_ageY} shows the differences in shapes and depths of three Balmer lines
in the integrated synthetic spectra of a GC computed using the method described in Section~\ref{method},
abundances of NGC\,5286 (Table~\ref{tab:abund}) and different isochrones by B08.
For this presentation, we select only four isochrones providing the best agreement between the synthetic and
the observed spectra. The synthetic Balmer lines computed with several other isochrones
($\rm log(Age)=10.15, Y=0.30, Z=0.0004$, $\rm log(Age)=10.15, Y=0.26, Z=0.0004$) are shown in 
Appendix~\ref{app_age_he}
(Fig.~\ref{fig_ageYadd}).
Figure~\ref{fig_ageYGCs} illustrates the comparison of the Balmer lines in the spectra
of NGC\,5286, 1904, 6254, 6752, and 7089.

Figs~\ref{fig_ageY} and \ref{fig_ageYadd} 
(Appendix~\ref{app_age_he}) show
that the cores and wings of the lines are fitted better in case 
 we choose the isochrone (B08): $\rm Z=0.0004, log(Age)=10.10, Y=0.30$.

Figure~\ref{fig_ageY} demonstrates that the depths of the Balmer lines in the presented synthetic 
spectra differ only by 1-2\% (Fig.~\ref{fig_ageY}), when we use four different isochrones
with the ages 12.6~Gyr and 11.2 Gyr and helium abundances Y$=0.26$ and 0.3.
The quality of observational data determines the uncertainties of age and Y.
It is worth mentioning that the Balmer lines are blended in the medium resolution spectra,
and different chemical elements (primarily Fe) re-shape their wings.
For example, Khamidullina et al. (2014) selected the best-fitting isochrone for NGC\,6254:
$\rm Z=0.0004, log(Age)=10.05, Y=0.30$ (B08). 
VandenBerg et al. (2013) 
determined for NGC\,6254: $\feh=-1.57$, Y$=0.25$ and the absolute age 11.75~Gyr. 
The last age estimate is between our result (Table~\ref{tab:prop5GCs}) 
and that of Khamidullina et al. (2014). 

\subsubsection{CMDs of five Galactic GCs in comparison with theoretical isochrones}
\label{compar_CMDs}
In this section, we will address the question: why
 the isochrone $\rm Z=0.0004, log(Age)=10.10, Y=0.30$ describes reasonably well the intensities 
 and shapes of Balmer lines in the medium-resolution integrated-light spectra of our sample Galactic GCs,
while it is known that our sample GCs possess multiple stellar populations,
and their HB morphologies can not be fully reproduced by an isochrone that is for a single Y value only.

Figure~\ref{figproportion} 
shows the CMDs of five Galactic GCs.
We used the deep CMD of NGC\,1904 from Piotto et al. (2002). 
Other high-quality photometric data were taken from the paper by Sarajedini et al. (2007).
Figure~\ref{figproportion}a illustrates our method of synthetic integrated-light spectra calculation 
(Section~\ref{method} and references therein) 
and compares the CMD of NGC\,1904 with the isochrone $\rm Z=0.0004, log(Age)=10.10, Y=0.30$ (B08).
One can see that the theoretical HB appears to be significantly more luminous than  
the level of the observed HB. The same situation takes place for other four GCs,
because the levels of the observational HBs look similar.
The structures of the observed HBs are complex
(e.g. Marino et al. 2015 and references therein, Schiavon et al. 2004).
They can not be exhaustively described by any theoretical isochrone.

Table~\ref{tab:proportion} presents the contribution of HB stars to the integrated light 
of the clusters in the B-band of the Johnson-Cousins photometric system
divided by the corresponding contribution of RGB stars brighter than the level of the HB.
The lower luminosity level chosen to select
RGB stars corresponds to the local maxima of luminosity on the RGB (B08).
Figs~\ref{figproportion}(b)--(e) illustrate the way how the stars belonging 
to the HB and upper RGB of GCs have been selected.
We used the B-band, because its wavelength coverage is close to the analysed 
spectral range (Section~\ref{method}).
We transformed stellar photometric data corrected for foreground extinction 
into the B-band as follows: $\rm (V-I)_0 = 0.893 (B-V)_0$. 
We obtained this formula for photometric standard stars (Landolt, 1992).

The values in the last four columns of 
Table~\ref{tab:proportion} represent the theoretical contribution of HB stars to the integrated
spectra of GCs computed using our method (Section~\ref{method} and references therein)
with respect to the corresponding contribution of upper RGB stars.
Theoretical contribution of a particular isochrone point to the total luminosity
of the cluster 
was calculated for the selected four isochrones from B08 using the mass function by Chabrier (2005).
Evolutionary stages (points of isochrones) 
with initial stellar masses differing by less than $10^{-4} M_{\sun}$ were excluded
from the calculations of synthetic integrated-light spectra of GCs.

One can conclude from Table~\ref{tab:proportion} that 
the contribution of HB stars to the integrated spectra of our sample GCs 
with respect to the corresponding contribution of RGB stars brighter than the level of HB
is reproduced well by the isochrone $\rm log(Age)=10.10, Y=0.30, Z=0.0004$ (B08).

To summarise, the
 integrated-light synthetic spectra calculated
using one isochrone can fit particularly well 
the observed integrated-light medium-resolution spectra of GCs, when the following conditions are satisfied: 
i) the range of stellar colours (effective temperatures) covered by the theoretical HB
agrees well with the corresponding observational data;
ii) contribution of the HB integrated light to the total luminosity of the cluster
is similar in the observational photometric and spectroscopic data, on the one hand, and in the 
integrated-light synthetic spectra calculated using this isochrone, on the other hand.

\section{Structure of the globular cluster in KKs\,3}
\label{structure}
\begin{figure}
\centering
\includegraphics[scale=0.36]{Fig6.ps}
\caption{F606W ACS image of the central GC in KKs\,3}
\label{figKingKKS3ima}
\includegraphics[angle=-90,scale=0.37]{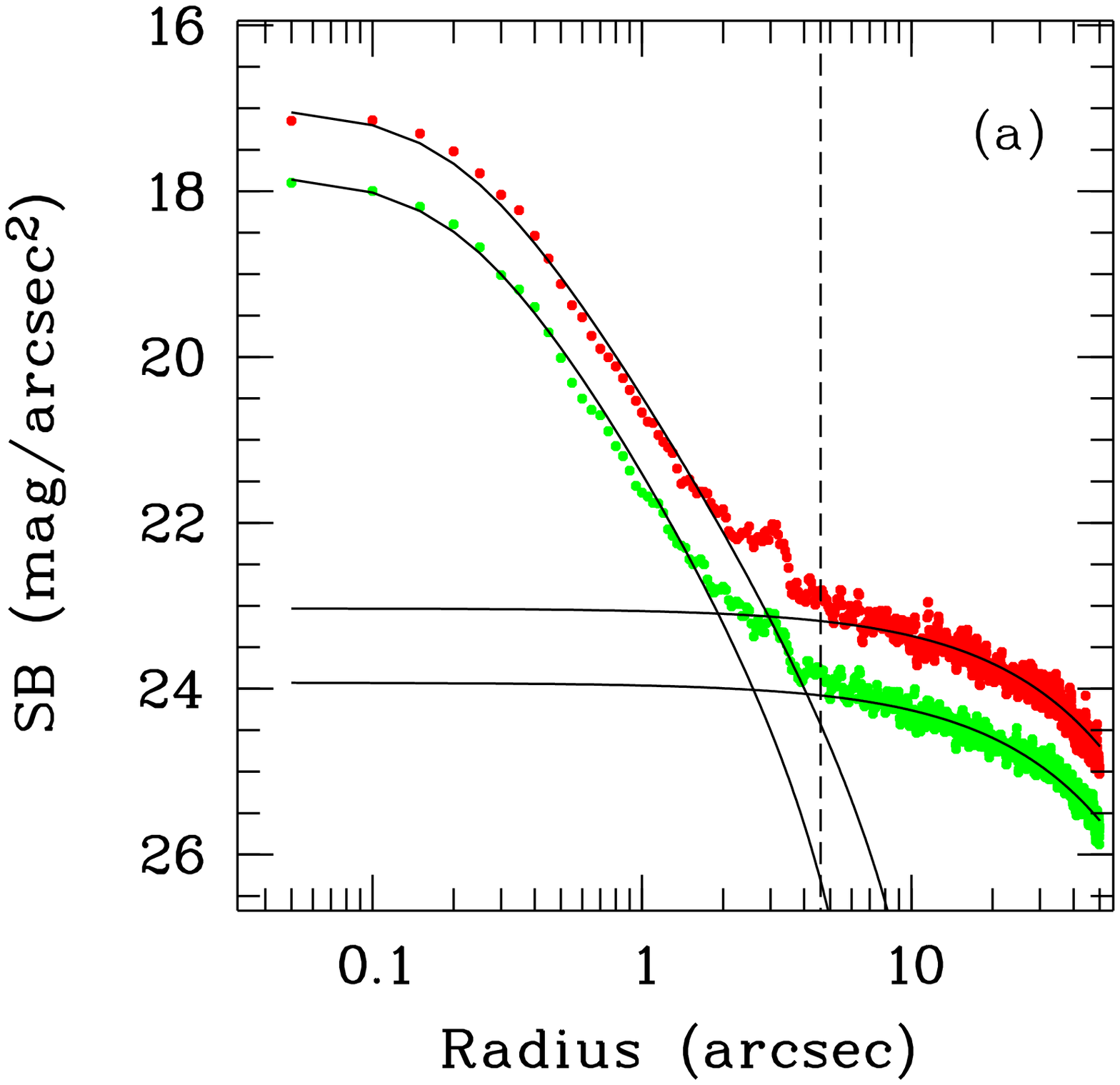}
\includegraphics[angle=-90,scale=0.37]{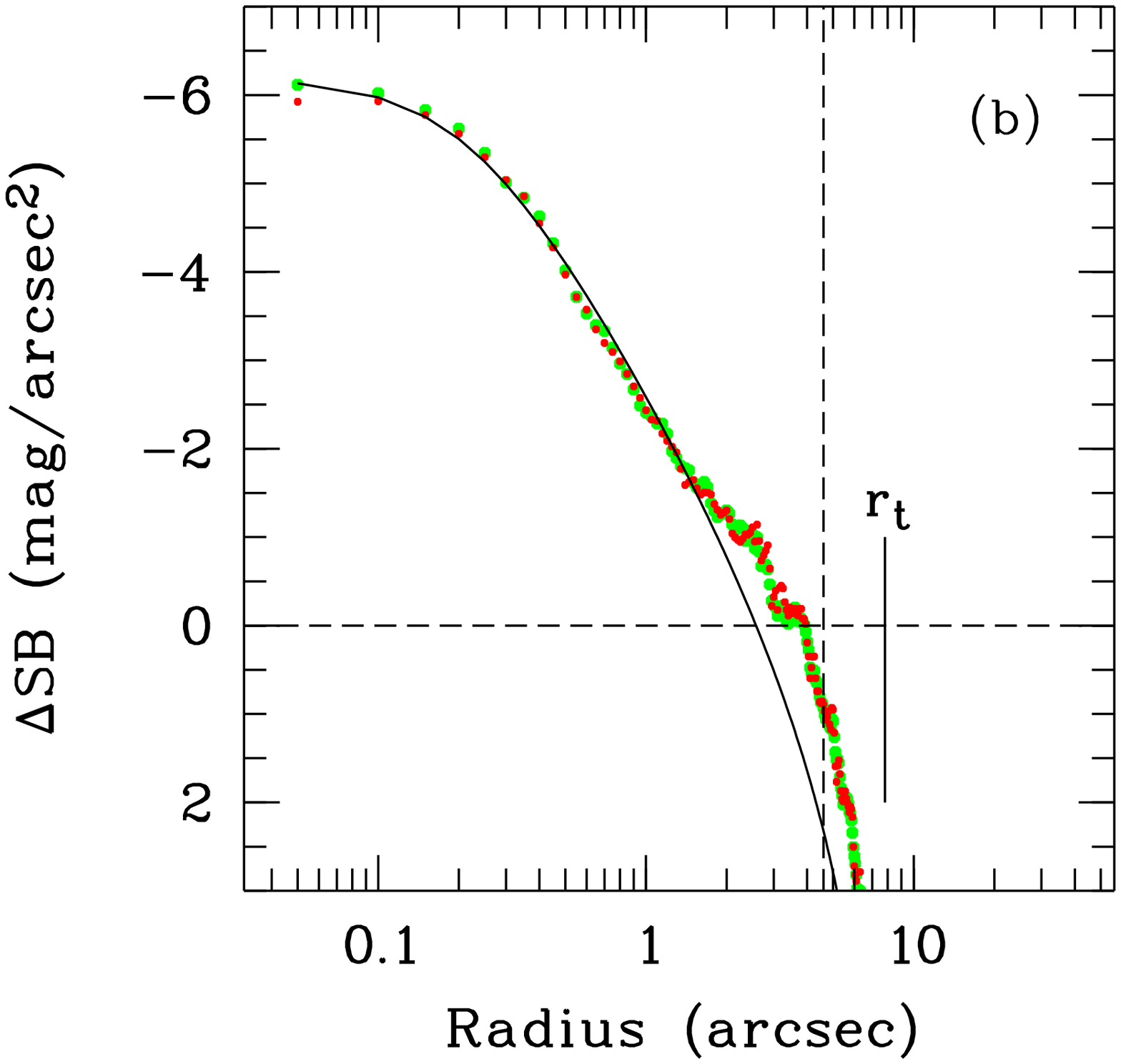}
\caption{(a): Azimuthally-averaged surface brightness profiles of the central region of KKs\,3 in the V and I bands (dots).
Solid lines show a decomposition of the profile inside the fit range represented by the exponential function 
and the King (1962) law.  (b): The same as in (a), but after subtraction of the galactic contribution.
The horizontal dashed line is a zero line. 
The vertical dashed line in panels (a) and (b) marks the radius ($\rm R\sim4.6\arcsec$) 
at which the galactic contribution begins to prevail.}
\label{figKingKKS3}
\end{figure}

Two 1200 s exposures obtained aboard the HST with the Advanced camera for surveys (ACS) 
and the filters F606W and F814W  
(SNAP program 13442) were used to derive structural parameters of the GC in KKs\,3.
 A fragment of the image containing the GC is shown in Fig.~\ref{figKingKKS3ima}.
The azimuthally-averaged surface brightness profiles of the central region of the galaxy 
including the GC was generated using aperture photometry with circular apertures of increasing size
on the F606W and F814W frames using the SURFPHOT program in the MIDAS package
 developed by ESO. Bright foreground stars 
 and background galaxies were masked. 
 The centre of the cluster was determined with an accuracy of $\sim$1 pixel
in each coordinate using the MIDAS routine FIT/ELL3. The background was defined by fitting a plane 
to the image with the FIT/BACKGROUND program.
Conversion from the instrumental photometric system to the V and I bands of the Johnson-Cousins
system was carried out using zeropoints
and calibration coefficients from Sirianni et al. (2005).

The integrated magnitudes of the cluster in the V and I bands were estimated to be 
$\rm V=18.41\pm0.03$ and $\rm I=17.45\pm0.05$ yielding the colour $\rm V-I=0.96\pm0.06$.
These results stay in good agreement with the estimates by Karachentsev et al. (2015b).
Please, note that the presence of a bright red star $\sim$9\arcsec\ to the South-West 
from the GC can influence surface photometry results.
The surface brightness profiles of the central region of KKs\,3 are shown in Fig.~\ref{figKingKKS3}(a).
The King (1962) profiles with the parameters listed in Table~\ref{tab:KingKKS3} the V and I bands 
are over-plotted. The models were fitted after subtraction of the galactic contribution.
The last one was estimated by fitting the exponential law to the azimuthally-averaged surface
brightness profiles with the parameters from Karachentsev et al. (2015a).

After subtraction of the galactic contribution, it becomes clear that 
there is brightness excess along the measured surface brightness profile 
at the radii between 2\arcsec and 4.6\arcsec 
reaching $\sim8\sigma$ above the background (Fig.~\ref{figKingKKS3}b), where $\sigma$ is
the dispersion of sky counts at the radii greater than $\rm r_t$ 
after subtraction of the galactic contribution: a mean $\sigma \sim 0.13$~mag.
The mentioned "extra" light constitutes $\sim10$\% of the total cluster luminosity integrated 
from the centre to the tidal radius. 

Our photometric results yield the half-light radius of the cluster $\rm r_h=4.8\pm0.2$~pc.
The object appears to be a bit more diffuse than the GC in ESO\,269-66 and than the
five Galactic GCs with similar spectra (Table~\ref{tab:prop5GCs}).
The concentration parameter $\rm c=log(r_t/r_c)= 1.6$ of the GC in KKs\,3 is typical 
for massive Galactic GCs (Harris, 1996).
\begin{table}
\caption{Structural parameters of GC in KKs\,3 measured using the the King (1962) law:
central surface brightness in mag~arcsec$^{-2}$, core and tidal radii in arcsec.} 
\label{tab:KingKKS3}
\begin{center}
\begin{tabular}{cccc}
\hline\hline
 $\mu^0_V$  & $\mu^0_I$  &  $\rm r_c$        & $\rm r_t$     \\  
\hline
 17.74$\pm$0.02 & 16.97$\pm$0.02 & 0.22$\pm$0.01 & 7.8$\pm$4.7 \\
\hline
\end{tabular}
\end{center}
\end{table}
\section{Discussion}
\label{Discussion}

\subsection{Multiple stellar populations in the nuclei of ESO\,269-66 and KKs\,3?}
\label{multStPops}
\begin{figure*}
\centering
\includegraphics[angle=-90,scale=0.58]{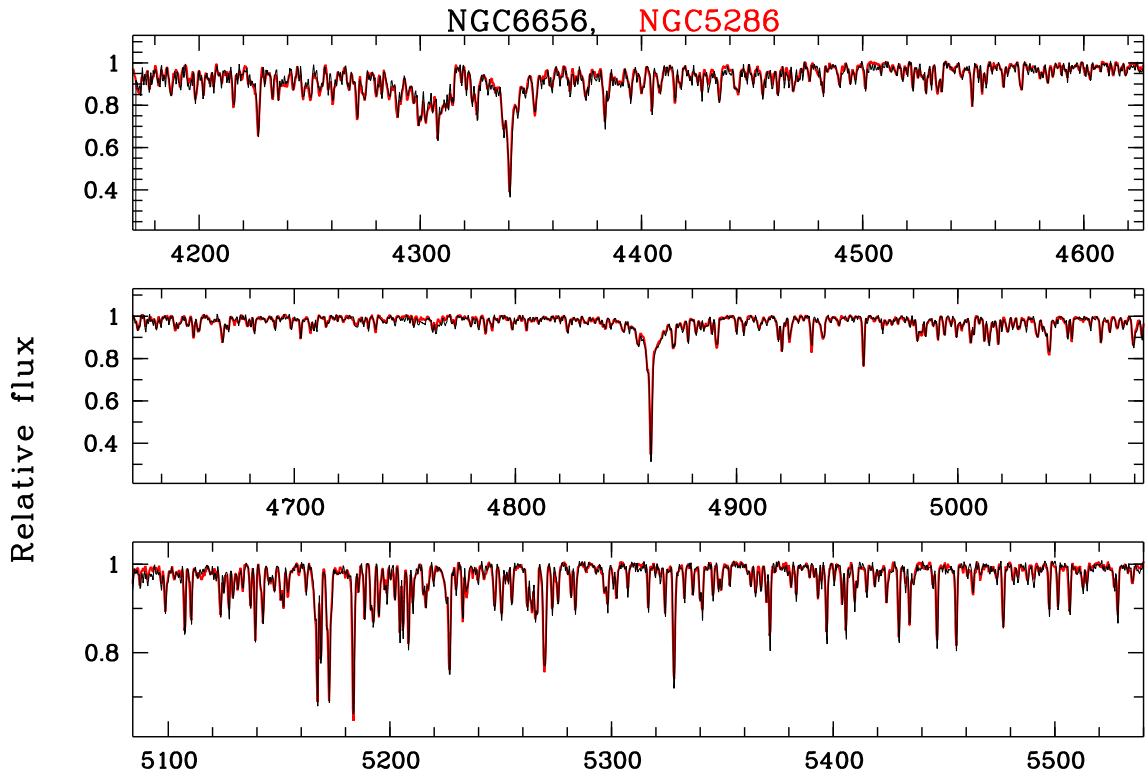}
\caption{Normalised spectra of NGC 5286 (red in the electronic edition) and NGC\,6656 (black) 
from WAGGS (Usher et al. 2017).}
\label{fig_N5286_6656_6715}%
\end{figure*}
As it was noted in Section~\ref{compar_abund}, according to the literature studies, all 
five Galactic counterparts of the two nuclei of dSphs exhibit the signs of multiple 
stellar populations in the sense of splits of their evolutionary sequences, correlated variations
of light-element abundances coupled with He enrichment and the appearance of EHBs populations.
NGC\,5286 and 7089 are anomalous GCs (Marino et al. 2015 and references therein). 
Can we find more analogues of the GCs in KKs\,3 and ESO\,269-66 using publicly available spectral data?

A new WiFeS Atlas of Galactic Globular cluster Spectra (WAGGS) (Usher et al. 2017)
provides 22 integrated-light spectra of GCs in the Milky Way satellites and of 64 old Galactic GCs.
This survey includes five objects of our study (NGC\,1904, 5286, 6254, 6752, 7089) and a dozen other GCs  
with $\feh\!\sim\!-1.6$~dex.
Two Large Magellanic Cloud GCs in this sample, NGC1786 and NGC1916, have similar metallicities and 
old ages (Usher et al. 2017 and references therein).
We compared the spectra of all these objects using the WAGGS spectrum of NGC\,5286 as a reference one. 
We used the original resolution ($\rm R=6800$) and 
all four gratings that cover the total spectral range: 3300 -- 9050\AA.
WAGGS  contains repeated observations for some GCs.
We used spectra having higher signal-to-noise ratio.
The results of the comparison are presented in Fig.~\ref{fig_N5286_6656_6715} 
and at an anonymous ftp website\footnote{
ftp://ftp.sao.ru/pub/sme/TwoDSphs/AppendixF.pdf}.
The comparison plots do not include the parts of the spectra distorted by strong telluric lines.

As a result, we found three additional GCs  with integrated-light spectra similar 
to those of the GCs in KKs\,3 and ESO\,269-66: NGC\,6656, 6273, and 6681. 
The depths and shapes of Balmer and FeI, II lines, molecular bands (CH, MgH, and CN), 
strong CaI, CaII, and MgI lines (CaII H+K, CaII 8498, 8542, 8662\AA\AA, CaI 4227\AA, and MgI 5183\AA), 
and of weaker lines and blends contributed by different chemical elements
look similar to the corresponding spectroscopic features in the WAGGS 
spectrum of NGC\,5286. 
On the other hand, NGC\,6254 looks a bit more metal-rich than NGC\,5286. 
The same trend was discovered in Section~\ref{results} (please, see also Larsen et al. 2017).
The spectrum of NGC\,5139 ($\omega$Cen) also demonstrates resemblance
to the spectrum of NGC\,5286. However, Fe lines look $\sim1\div2$\% less intensive in the spectrum
of NGC\,5139. Johnson et al. (2017) reported on the close identity of Fe I and Fe II line profiles 
of stars in three metallicity groups belonging to NGC\,5139 and NGC\,6723 with similar 
$\rm T_{eff}$, $\rm log(g)$, and \feh.
We suggest that determination of ages and Y and accurate abundance analysis for the found
GCs with similar spectra is a subject of a separate study.

To summarise, 
 all five known Fe-complex Galactic GCs with $\feh\sim-1.6$~dex 
(Marino et al. 2015, Johnson et al. 2017) have medium-resolution spectra resembling the spectra 
of GCs in KKs\,3 and ESO\,269-66: NGC\,5286, 7089, 6656, 6273 and 5139. 
All nine possible Galactic analogues of the GCs in KKs\,3 and ESO\,269-66 demonstrate the signs 
of multiple stellar populations  (Piotto et al. 2015; Milone et al. 2013, 2015; Marino et al. 2015; 
Carretta et al. 2009, 2010; Fabbian et al. 2005; O'Malley et al. 2017). 
Please note, that the discovered resemblance of the spectra is surprising, because 
integrated-light spectra can be influenced by the Galactic field contamination and by 
effects of stochastic fluctuations in a number 
of stars within the spectrograph field-of-view.

\begin{table}                                                                                                                             
\caption{Properties of the selected Galactic GCs (please see text for details). Fe-complex GCs
(Marino et al. 2015, Johnson et al. 2017) are marked by asterisk.} 
\label{tab:prop9GCs}
\scriptsize
\begin{center}
\begin{tabular}{lccccrccc}
\hline\hline                                                                                                                       
NGC    & $\rm D_{MW}$  & Z     &  E(B-V)       & $\rm M_{V_0}$&HBR   &$\rm R_h$&$\feh$   \\ 
       &  (2)          &  (3)  &     (4)       & (5)          & (6)  & (7)     & (8)           \\
\hline                                                                                   
1904   &   18.8        & -6.3  &    0.01       & -8.23        &0.89  & 3.00    &-1.58        \\
5286$^*$&   8.4        & 2.0   &    0.24       & -9.45        &0.80  &  2.21   &-1.70        \\
6254   &   4.6         & 1.7   &    0.28       & -8.34        &0.98  & 2.32    &-1.57        \\
6752   &   5.2         & -1.7  &    0.04       & -7.85        &1.00  & 2.72    &-1.55        \\
7089$^*$&  10.4        & -6.7  &    0.06       & -9.21        &0.96  & 3.11    &-1.66        \\
\noalign{\smallskip}
5139$^*$&  6.4         &  1.4  &    0.12       & -10.63       & --   & 6.44    &-1.64        \\
6273$^*$&  1.6          & 1.4   &    0.38       & -10.29       & --   & 3.13    &-1.76        \\
6656$^*$&  4.9         & -0.4  &    0.34       & -9.54        &0.91  & 3.03    &-1.70        \\
6681   &  2.1          & -1.9  &    0.07       & -7.33        &0.96  & 2.43    &-1.62        \\
\hline                                                                                                                                                 
\end{tabular}
\end{center}
\end{table}
Properties of the objects with the spectra similar to the 
spectra of KKs\,3 and ESO\,269-66 are summarised in Table~\ref{tab:prop9GCs}.
The columns are the following: 
(2-6) distance from the Galactic centre (kpc), 
Z distance component toward the North Galactic pole (kpc), 
colour excess (mag), absolute magnitude (mag), and                                      
HB ratio ($\rm HBR = (B-R)/(B+V+R)$) from Harris (1996), 
(7) half-light radius (pc) from Mackey \& van den Bergh (2005),             
(8) \feh\ (dex) from Carretta et al. (2010 and references therein).
One may conclude that 
the listed objects are among the brightest GCs in our Galaxy.
They have low metallicity $\feh\!\sim\!-1.6$~dex  
and old age (e.g. Caretta et al. 2010, VandenBerg et al. 2013).
All the GCs, except for $\omega$Cen, have similar half-light radii $\rm R_h=2.7\pm0.3$~pc 
and extended blue HBs (Piotto et al. 2015).

The objects listed in Table~\ref{tab:prop9GCs} do not form an ensemble belonging to any 
known Galactic tidal stream (Grillmair \& Carlin 2016), although some of them
are thought to be associated with streams (e.g. Bellazzini et al. 2003, Penarrubia et al. 2005).
Low Galactic latitudes and Z distance components of the GCs 
may indicate that they have participated in the process of the Galactic thick disk build-up.
Six of the sample GCs were classified as disk/bulge objects by Carretta et al. (2010) basing on 
observations of thousands of stars in GCs.
The thick disk is composed mostly of old stars and dominates the stellar number density 
between one and five kpc above the Galactic plane (Gilmore \& Reid 1983). 
Future high-resolution observations of many stars with 
different masses, $\rm T_{eff}$ and $\rm log(g)$ and
accurate comparison of chemical abundances of stars in the sample GCs, dSphs, and the thick Galactic disk
will help to make conclusions concerning the origin of these cosmological structures.

\subsection{On the origin of KKs\,3 and of its nucleus}

 Early-type dwarf galaxies are usually situated in close vicinity of giant neighbours.
Their morphologies are strongly influenced by environmental processes.
What is then the origin of KKs\,3 and of its nucleus? 
KKs\,3 is located far from nearby massive galaxies and from the Local Void
whose expansion does not affect the motion of the galaxy (Karachentsev et al. 2015a).
If KKs\,3 is not a runaway object from the Local group, it is probably a remnant of collision of 
two or three small galaxies (e.g. V{\"a}is{\"a}nen et al. 2014).
Pairs and triplets of small galaxies are found in different environments 
(e.g. Argudo-Fernandez et al. 2015 and references therein).
Three star forming episodes in KKs\,3 might be the evidences of these events.
Katkov, Kniazev \& Sil'chenko (2015) considered ionized gas accretion and satellite merging as the most probable 
explanation for the results of their studies of kinematics and metallicity of gas
in several isolated S0s.
Objects with small old or active nuclei are ubiquitous among low mass spiral galaxies.
If KKs\,3 is a low-mass analogue of S0s (Kormendy \& Bender 2012), its occurrence in a very
rarefied environment is not surprising. There is a large number of brighter lenticular
galaxies in the field (e.g. Sulentic et al. 2006).
The mass of gas involving in a collision event determines the duration and 
the intensity of star formation in the galactic centre.
Observations show that galactic nuclear activity is a main driver of morphological 
transformation of intermediate-mass early-type galaxies in any environments  
and that the galactic surface mass
densities, concentration indexes, and luminosities of galactic nuclei grow with 
the corresponding galaxy mass 
(Kauffmann et al. 2003, Argudo-Fernandez et al. 2016). 

There are many details to be added on, but we consider this issue to be a matter of a separate study.

\section{Concluding remarks}

In this paper, we analysed properties of stellar populations of 
nuclear star clusters in two dwarf spheroidal galaxies situated
in very different environments:
a unique isolated dSph KKS3 discovered by Karachentsev et al. (2015b) and ESO\,269-66, 
a close dSph neighbour of Cen~A (Karachentsev et al. 2007).
The galaxies have similar star formation histories and mean stellar metallicities.

We estimated ages, helium abundance (Y), \feh\ , and abundances of C, N, Ca, Mg, Ti, and Cr for
the nuclei using their medium-resolution spectra acquired with the Robert Stobie Spectrograph at 
the Southern African Large Telescope. To perform this work, we employed our developing method 
(Sharina et al. 2013, Khamidullina et al. 2014, Sharina et al. 2014) of computation
of integrated-light synthetic spectra of GCs using models of stellar atmospheres
according to the mass function by Chabrier (2005) and stellar parameters defined by
the selected isochrone from the paper by B08. 
The derived age of 12.6~Gyr, Y=$0.3$, $\feh \sim 1.5 \div 1.55$~dex and abundances of
chemical elements appear to be in the agreement with the 
corresponding data for five Galactic GCs from the sample of Sch05:
NGC\,1904, 5286, 6254, 6752, and 7089. Ages and elemental abundances for the Galactic GCs 
were estimated using our method and medium-resolution spectra from the library of Sch05.
The average abundances of chemical elements from the high-resolution literature spectroscopic 
studies for the considered Galactic GCs confirm our findings concerning the similarity.

Additionally, we found four objects in WAGGS (Usher et al. 2017) 
with integrated-light spectra resembling the WAGGS spectra of the
five Galactic counterparts of the GCs in KKs\,3 and ESO\,269-66. The objects are: 
NGC\,6273, 6656, 6681, and 5139. 
Determination of ages and Y and accurate abundance analysis for these GCs 
is a subject of a separate study.

Eight of the selected Galactic analogues
of the GCs in KKS3 and ESO\,269-66 (Table~\ref{tab:prop9GCs}),
posses extended blue HBs, correlated variations of light elements,
and splits of evolutionary sequences (Piotto et al. 2015, Milone et al. 2013, 2015; Marino et al. 2015, Carretta et
al. 2010, Fabbian et al. 2005, O'Malley et al. 2017). 
The ninth GC with the similar WAGGS spectrum, NGC\,6681,
overlaps the main body of Sagittarius dSph. 
Five of the found counterparts are iron-complex GCs:
NGC\,5286, 7089, 6273, 6656, and 5139 (Marino et al. 2015 and references therein, Johnson et al. 2017).

All the selected Galactic analogues of our two nuclei are at the extreme high end of the GC mass function.
All of them, except NGC\,5139 ($\omega$Cen), have similar half-light radii ($R_h=2.7\pm0.3$). 
The GC in ESO\,269-66 is exactly of the same size (Georgiev et al. 2009).
The GC in KKs\,3 is
slightly more extended ($\rm r_h=4.8\pm0.2$~pc) according to our measurements using HST images.
There is brightness excess near the end of the surface brightness profile of this cluster
which is not described by the King (1962) law. This "extra"-light constitutes $\sim10$\% of the total
cluster luminosity.

One may conclude that our method described in Section~\ref{method} is capable to select 
GCs with EHBs, most of which are very massive and some of which belong to a rare class of
anomalous GCs.
It is intriguing that GCs as the witnesses of most powerful star 
forming episodes in galaxies show similar properties of stellar populations 
independent on the environments.


\section{Acknowledgements}
We thank the anonymous referee for comments that helped to improve the paper.
All spectroscopic observations reported in this paper were obtained with the Southern
African Large Telescope (SALT) under programmes \mbox{2010-1-RSA\_OTH-002} and
\mbox{2014-2-MLT-001} (PI: Kniazev). 
The work is performed under the support of the grant of the Russian Science Foundation 14-12-00965
and in compliance with the Russian Government Program of Competitive Growth of Kazan Federal University.
A.\,K. acknowledges support from the National Research Foundation (NRF) of South
Africa and from the Russian Science Foundation (project no. 14-50-00043).
We thank L.N.~Makarova for providing us stellar photometric data for KKs\,3.
The work is based on observations made with the NASA/ESA Hubble Space
Telescope. STScI is operated by the Association of Universities 
for Research in Astronomy, Inc. under NASA contract NAS 526555.

\clearpage
\appendix
\onecolumn
\clearpage
\section{Determination of ages and helium abundances using integrated spectra of GCs}
\label{app_age_he}
\begin{figure*}
\hspace{-0.5cm}
\centering
\includegraphics[angle=-90,scale=0.64]{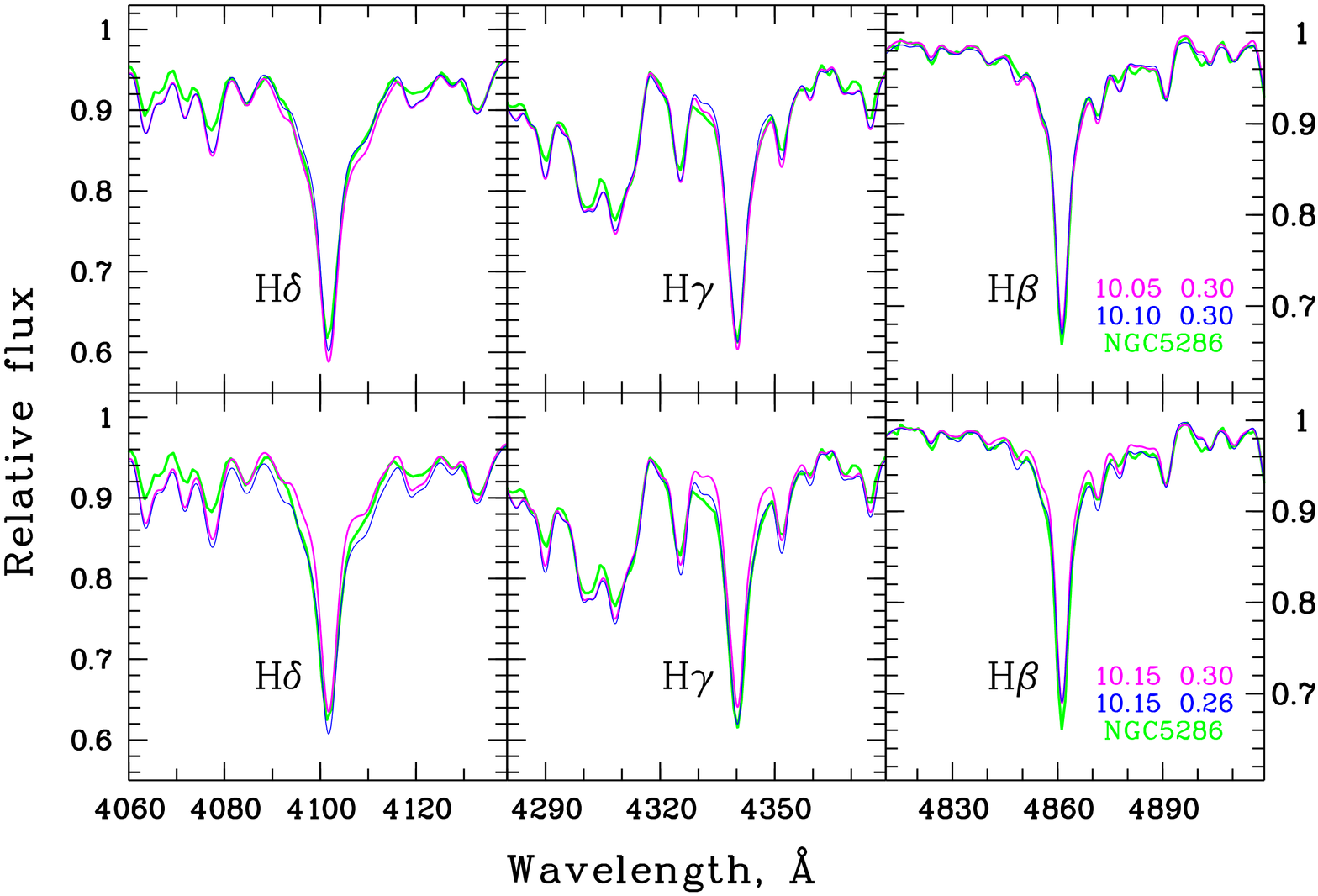}
\caption{Comparison of the spectrum of NGC\,5286 and computed synthetic integrated-light spectra of GCs 
in the ranges of the three Balmer lines. We used for the calculation the same elemental abundances valid for NGC\,5286
(Table~\ref{tab:abund}) and isochrones by Bertelli et al. 2008 of the same metallicity $Z=0.0004$ and 
different age and specific helium abundance: 
$\rm log(Age)=10.15, Y=0.30$, $\rm log(Age)=10.15, Y=0.26$, $\rm log(Age)=10.05, Y=0.30$ and $\rm log(Age)=10.10, Y=0.30$.}
\label{fig_ageYadd}%
\end{figure*}

\begin{figure*}
\hspace{-1cm}
\centering
\includegraphics[angle=-90,scale=0.6]{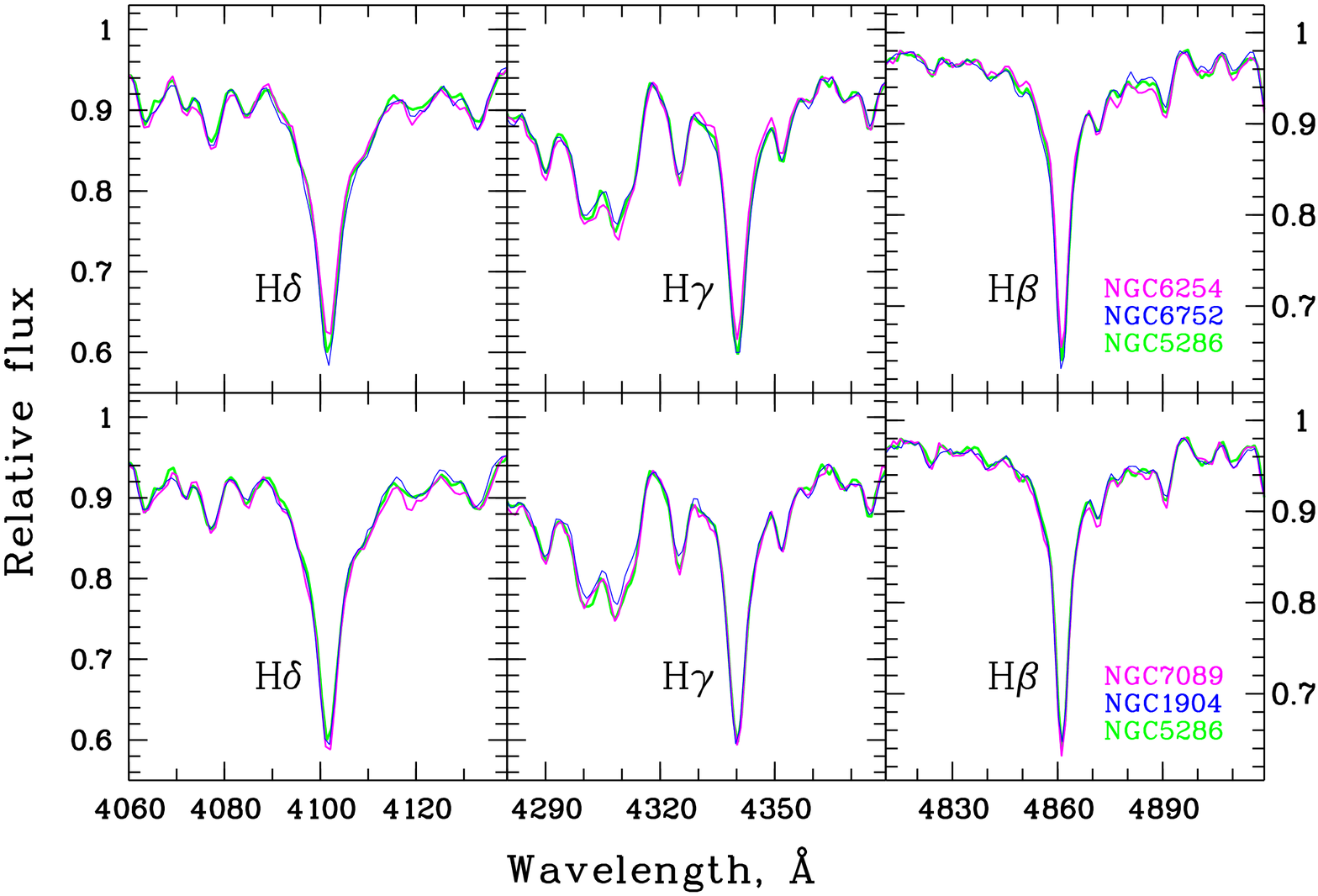}
\caption{Comparison of the three Balmer lines in the spectra of NGC\,5286, 1904, 6254, 6752 and 7089 .}
\label{fig_ageYGCs}%
\end{figure*}

Figures of this section illustrate our method of determination of ages and mean helium abundance (Y)
using integrated spectra of GCs (Section~\ref{method} and references therein). 
Figure~\ref{fig_ageYadd} compares the spectrum of NGC\,5286 (green line) from Schiavon et al. (2005) 
with the synthetic spectra computed using four 
different isochrones by Bertelli et al. (2008). It can be seen that no isochrone describes
exhaustively all the Balmer lines. The most suitable isochrone for fitting the spectrum of NGC\,5286 is:
$Z=0.0004$, $\rm log(Age)=10.10, Y=0.30$. Fitting the spectrum of NGC\,5286 with this isochrone
is shown in Fig.~\ref{fig_ageY} (Section~\ref{method}). Figure~\ref{fig_ageYGCs} illustrates the comparison of the Balmer
lines in the spectra of NGC 5286, 1904, 6254, 6752, and 7089. It can be seen that there is no exact
coincidence between the Balmer lines in five spectra. The aforementioned dissimilarities can be attributed to the 
complexities of the HBs, differences in the signal-to-noise ratios in the spectra, Galactic field contamination and 
effects of stochastic fluctuations in a number of stars within the spectrograph field-of-view
\clearpage

\section{Determination of microturbulent velocity and fitting the abundances of different chemical elements}
\label{app_elements}
\begin{figure}
\centering
\includegraphics[angle=-90,scale=0.40]{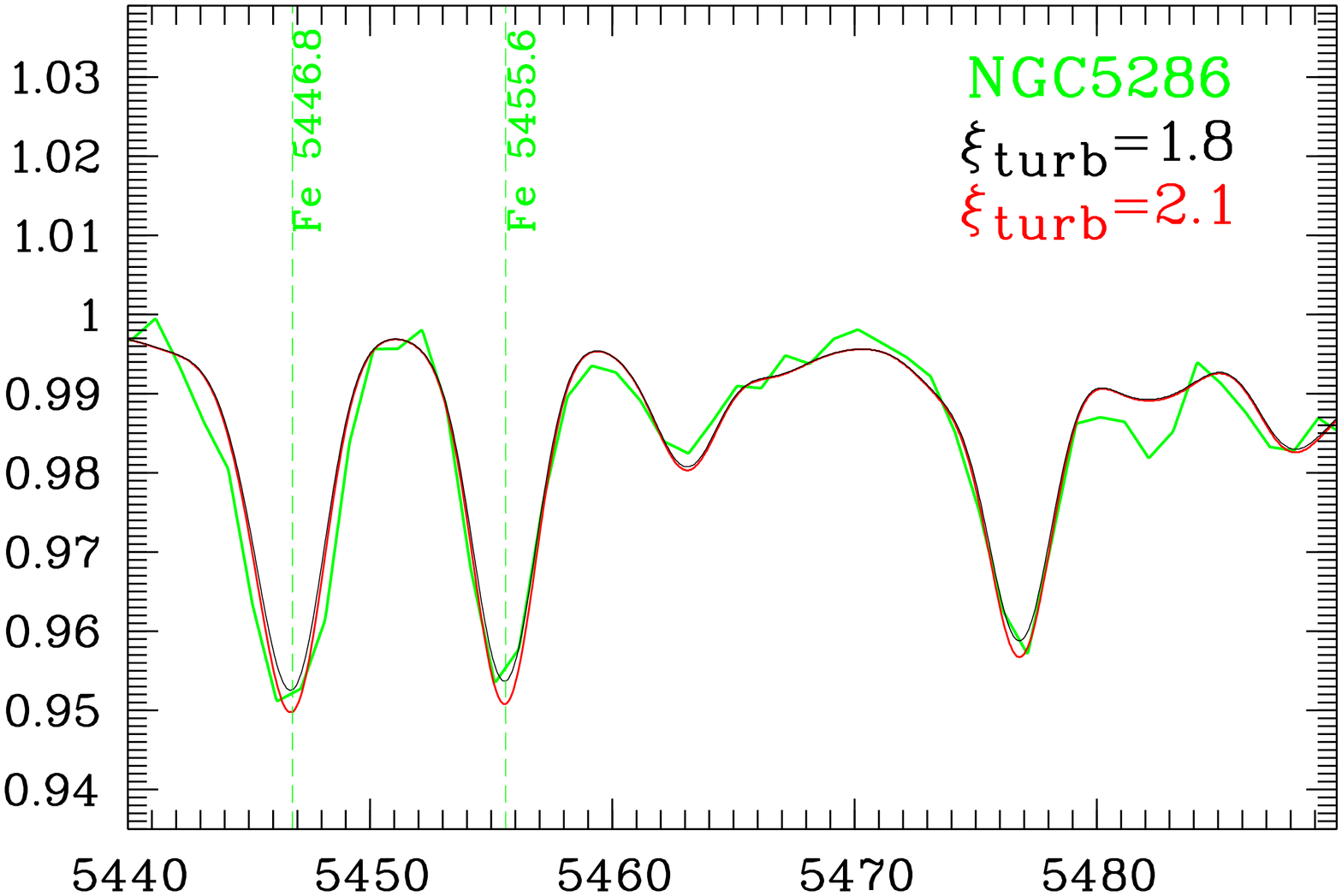}
\includegraphics[angle=-90,scale=0.40]{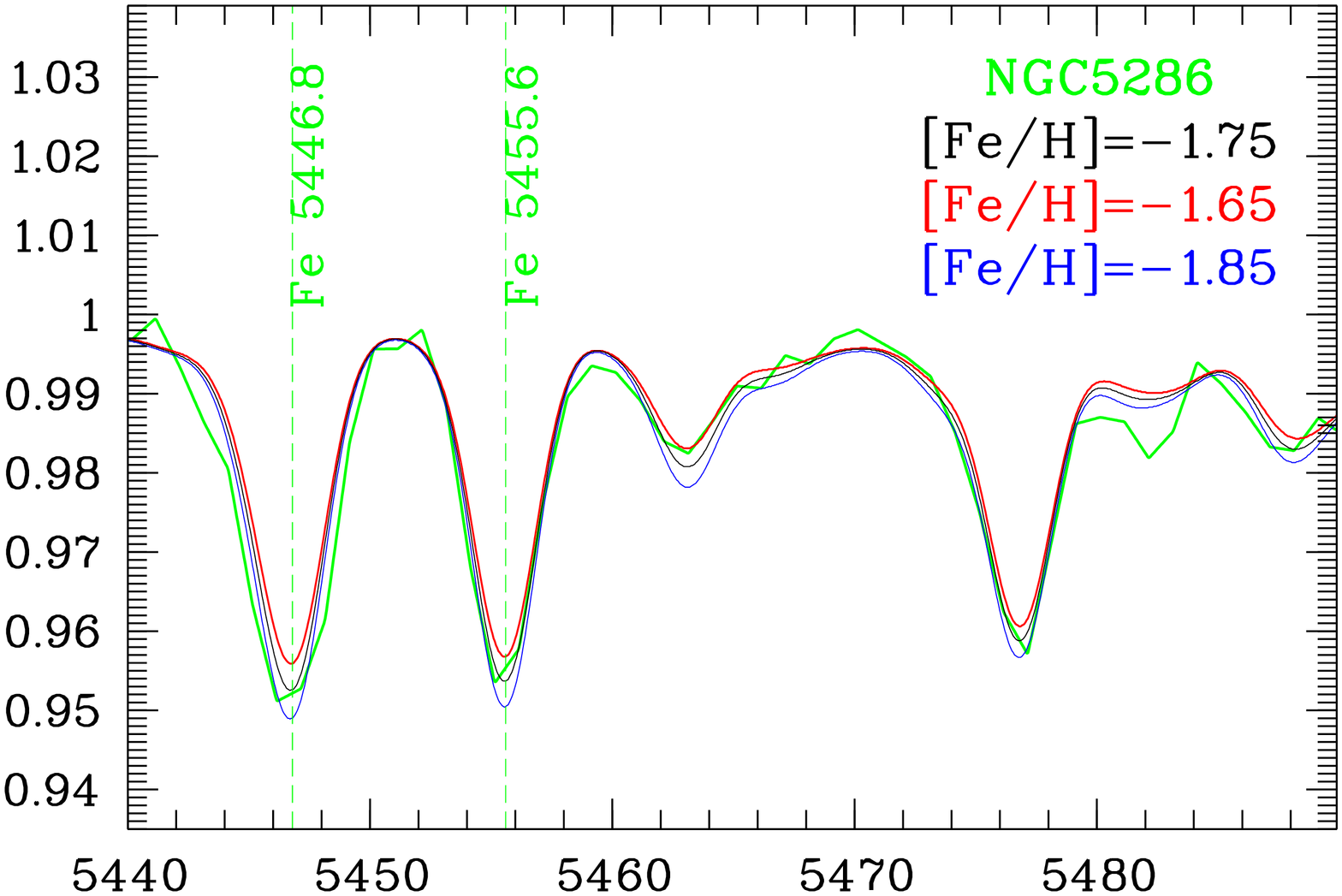}
\caption{Spectrum of NGC\,5286 (green line) in comparison to the synthetic spectra
computed with different $\xi_{turb}$ parameters (top) and different Fe abundances 
(bottom) and fixed other parameters. Relative flux is shown along the vertical axis.
Wavelengths are along the horizontal axis.}
\label{fig_XiTurbFe}%
\end{figure}

\begin{figure}
\centering
\includegraphics[angle=-90,scale=0.40]{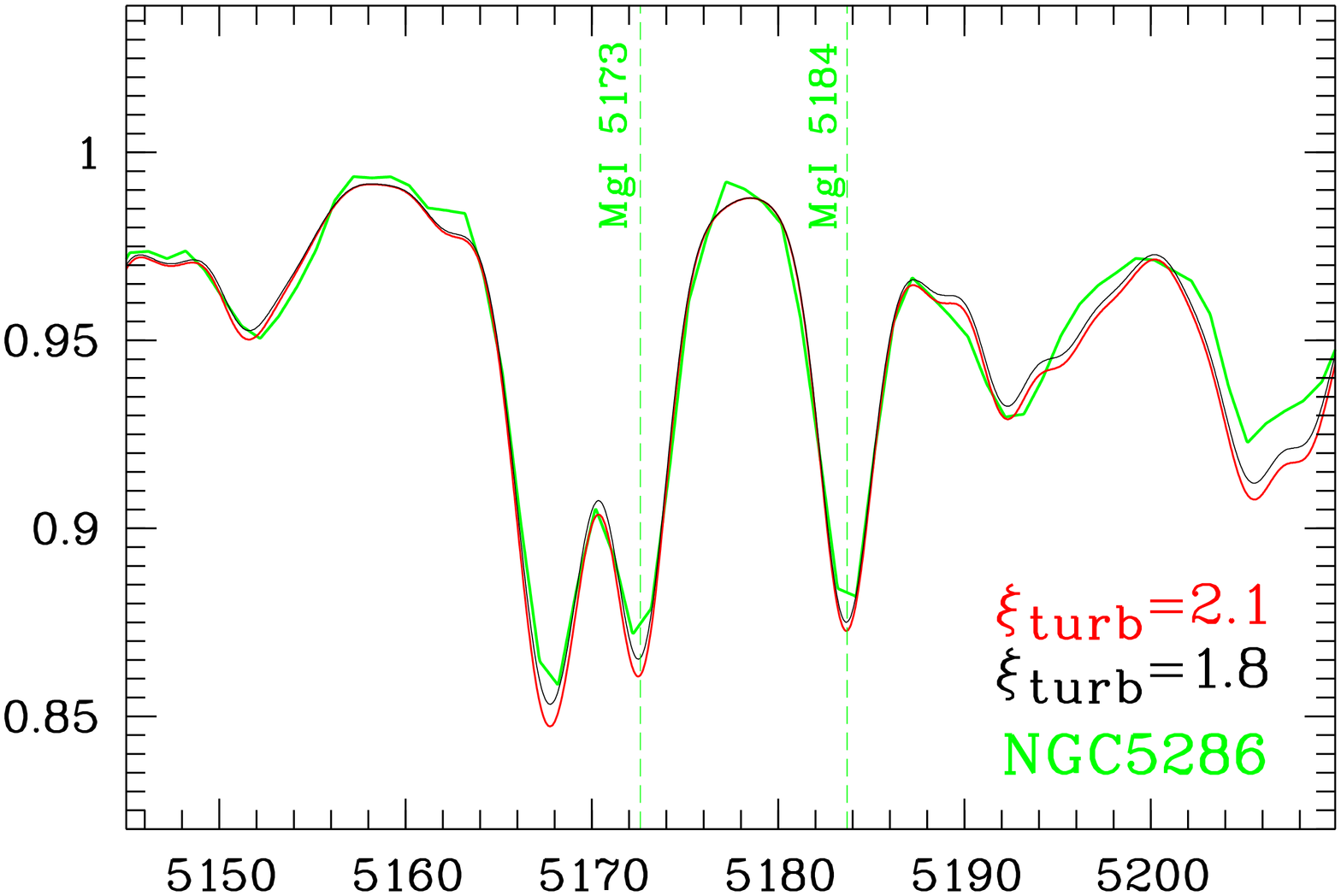}
\includegraphics[angle=-90,scale=0.40]{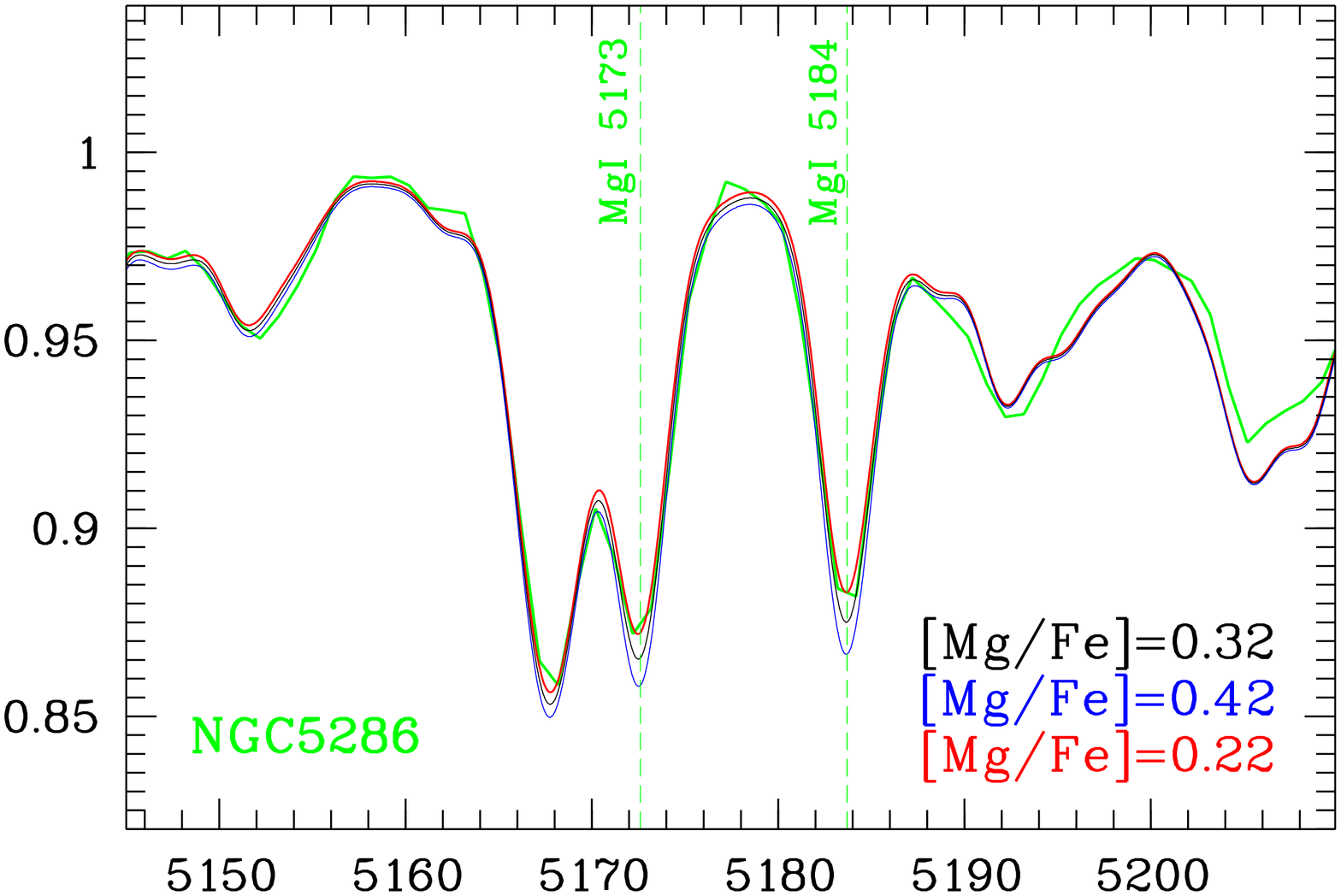}
\caption{The same as in Fig.~\ref{fig_XiTurbFe}, but for Mg abundances.}
\label{fig_XiTurbMg}%
\end{figure}

This section illustrates the process of determination of Fe, Mg, Ca, C, and Cr abundances
using medium-resolution integrated spectra of GCs. We use the spectrum of NGC\,5286 (Schiavon et al. 2005) as an example.
The method of chemical abundances determination is described in Section~\ref{method}. 
The results are shown in Table~\ref{tab:abund} and described in Section~\ref{results}.
Figs~\ref{fig_XiTurbFe} and \ref{fig_XiTurbMg} illustrate the influence of micro-turbulent velocity $\xi_{turf}$
on the depths of Fe and Mg line blends in the synthetic spectra of GCs.
\label{abundPlots}

\begin{figure}
\centering
\includegraphics[angle=-90,scale=0.40]{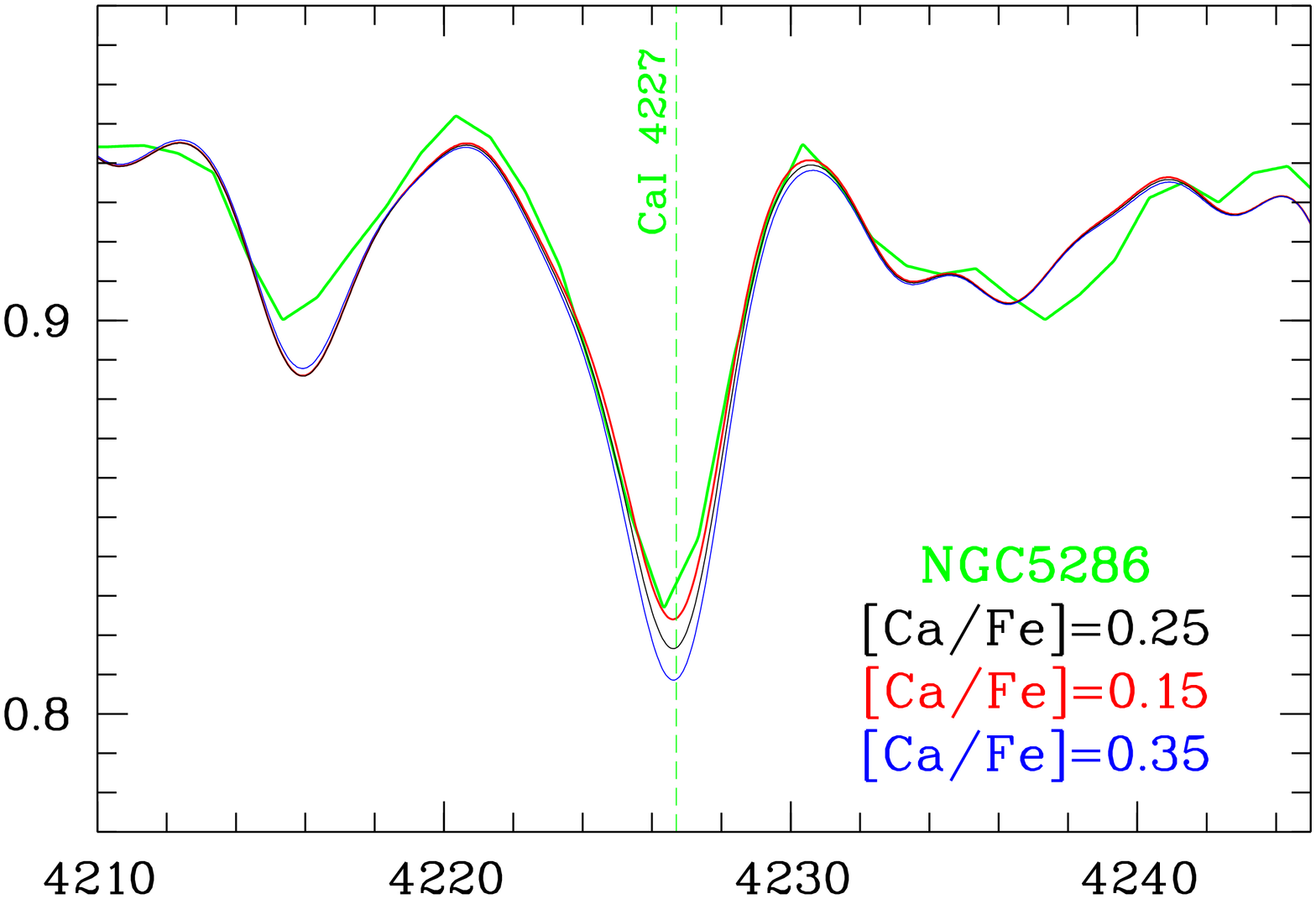}
\includegraphics[angle=-90,scale=0.40]{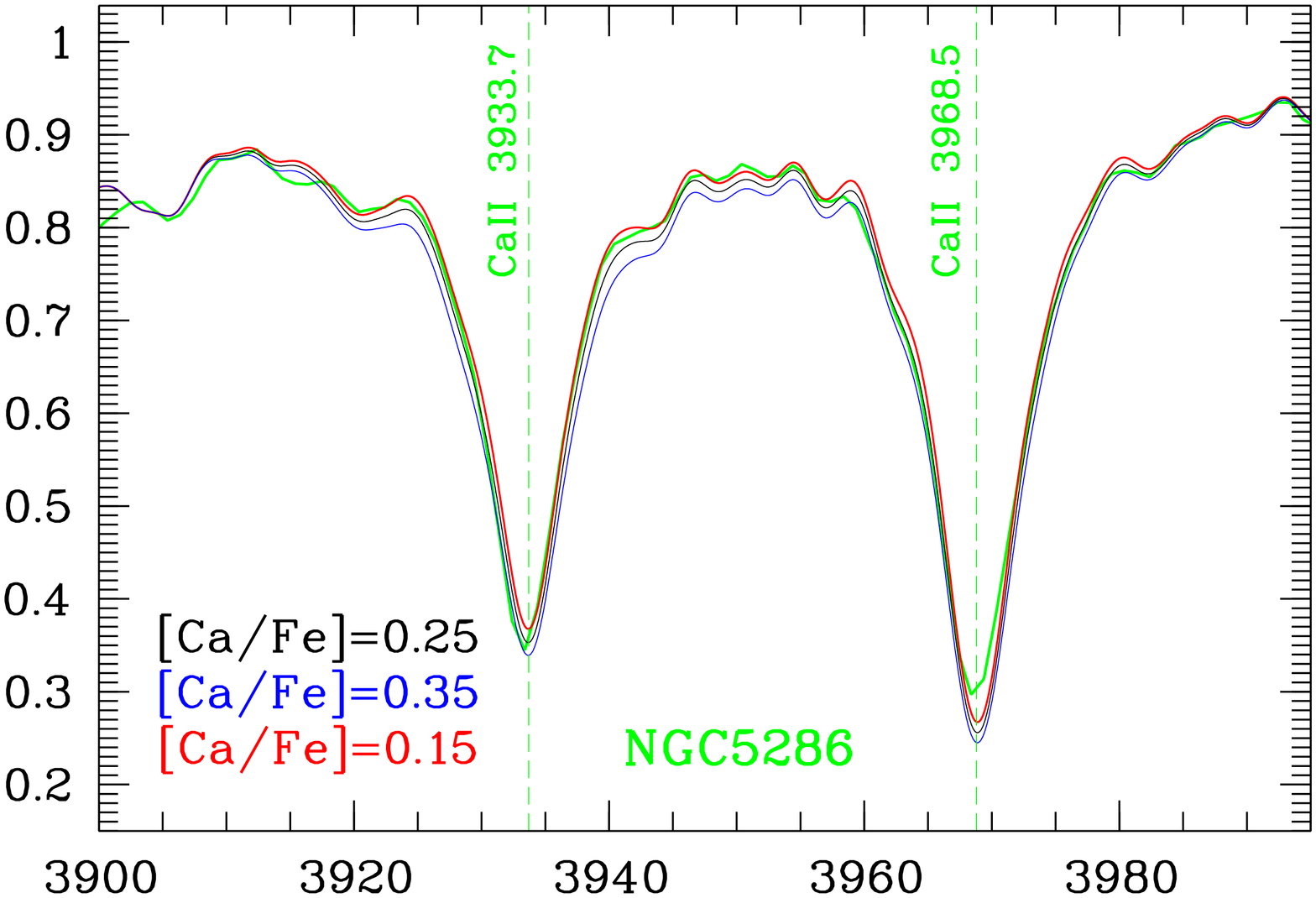}
\caption{Spectrum of NGC\,5286 (green line) in comparison to the synthetic spectra
computed with different Ca abundance and fixed other parameters.
Relative flux is shown along the vertical axis.
Wavelengths are along the horizontal axis.
}
\label{fig_Ca}%
\end{figure}
\begin{figure}
\centering
\includegraphics[angle=-90,scale=0.40]{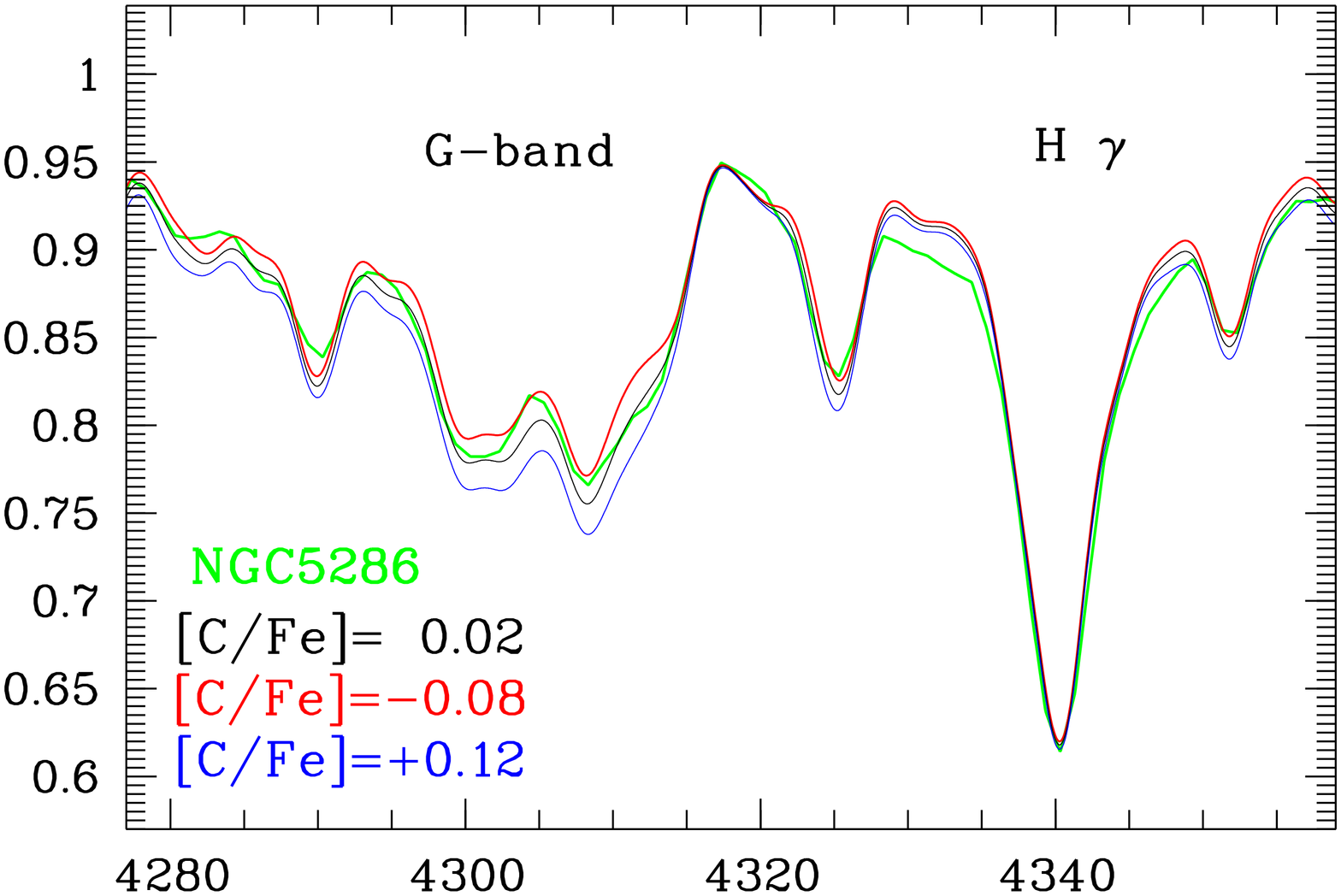}
\includegraphics[angle=-90,scale=0.40]{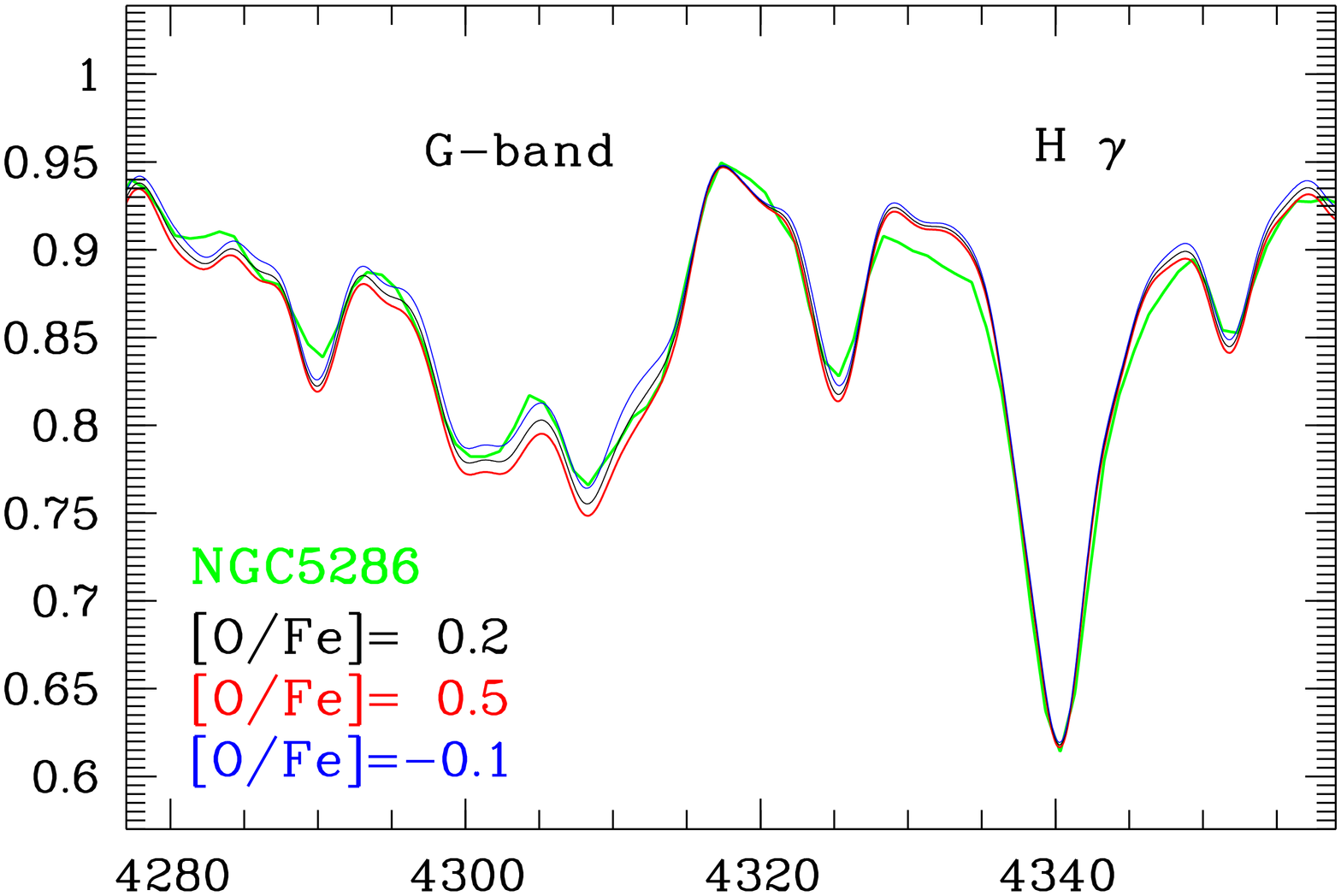}
\caption{The same as in Fig.~\ref{fig_Ca}, but for C and O abundances.}
\label{fig_CO}%
\end{figure}
\begin{figure}
\centering
\includegraphics[angle=-90,scale=0.40]{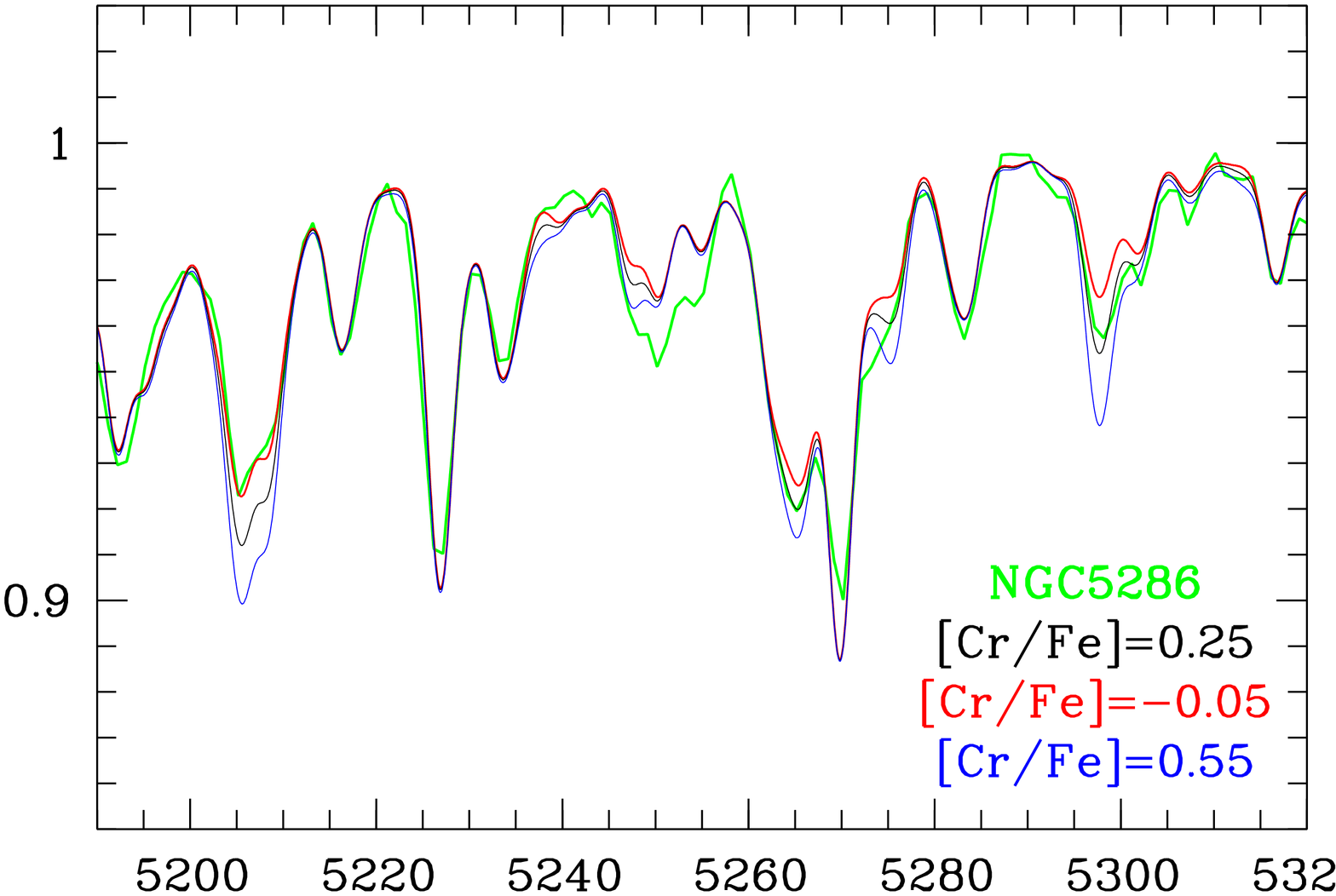}
\caption{The same as in Fig.~\ref{fig_Ca}, but for Cr abundance.}
\label{fig_Cr}%
\end{figure}

\clearpage
\section{Comparison of stellar photometric data for KKs\,3, ESO\,269-66, and for their nuclei}
\label{StelPhot}

\begin{figure}
\hspace{0.25cm}
\includegraphics[angle=-90,scale=0.305]{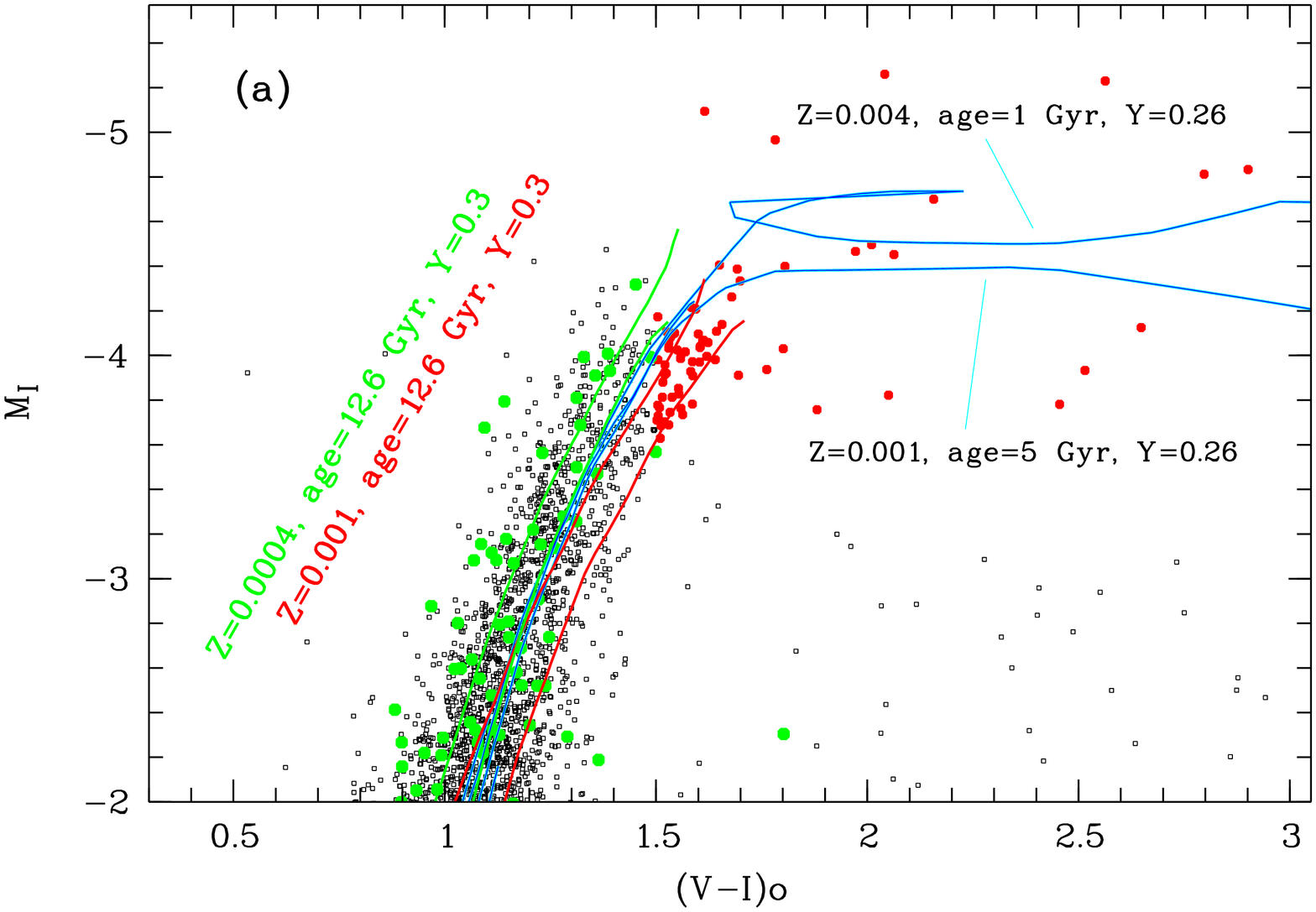}
\includegraphics[angle=-90,scale=0.32]{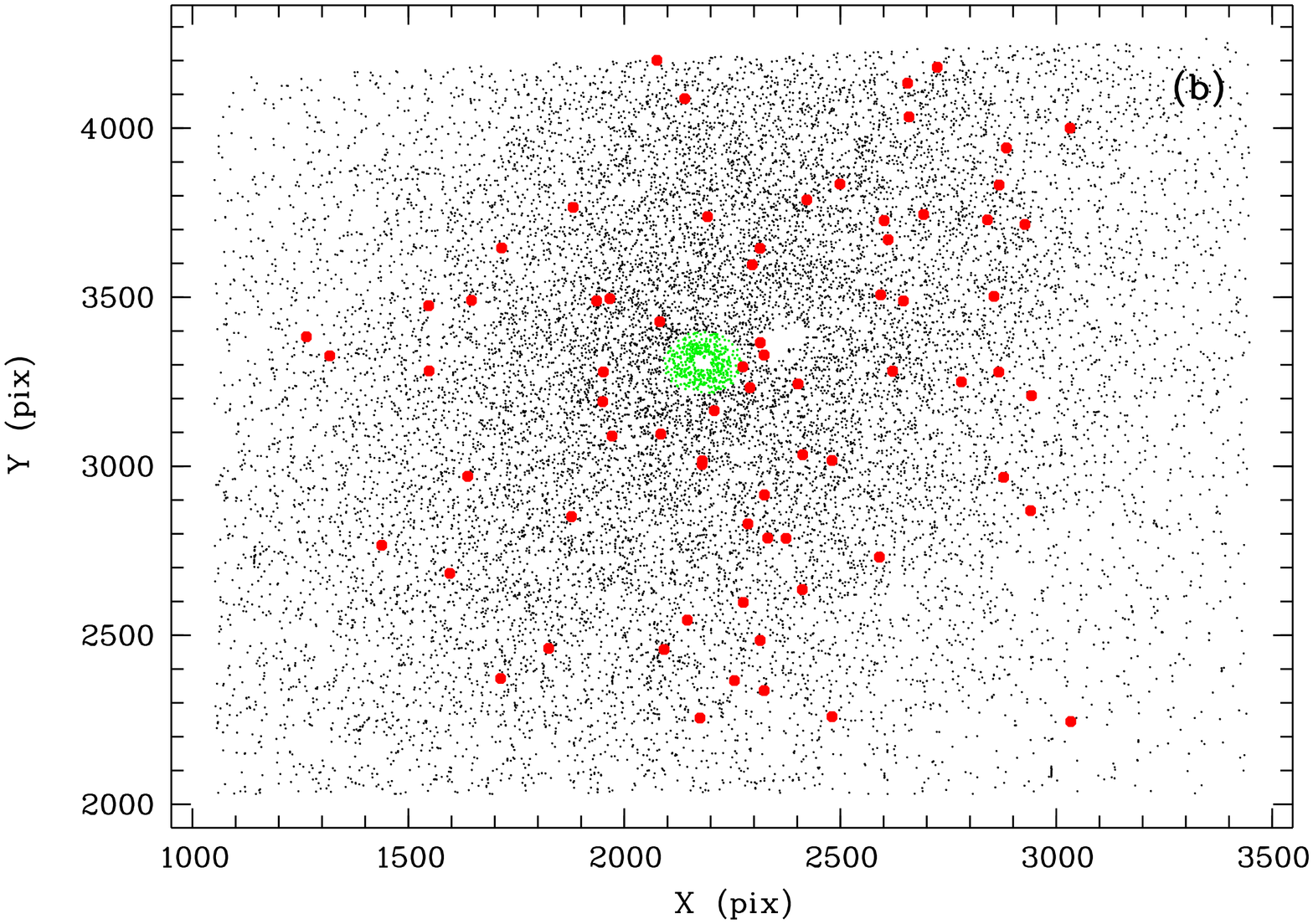}
\caption{(a) CMD for a central region ($\sim1\times1$~kpc) of KKs\,3. Stars within the radius of 47~pc 
around the centre of the GCs are shown in green. 
Bright red stars are shown as large red dots.
Four isochrones (Bertelli et al. 2009) are overplotted. 
(b) Location of the stars shown in the panel (a) within the HST frame.}
\label{fig:StelPhotKKS3}%
\end{figure}

High-quality photometric results for KKs\,3 were kindly provided by L.N.~Makarova.
Please, see the paper by Karachentsev et al. (2015b) for description of the data.
Transformation of the stellar magnitudes from the instrumental F606W and F814W filters
into the V and I bands of the Johnson-Cousins system was performed using the DOLPHOT package (Dolphin 2002).
Objects with large photometric errors, or non-stellar shapes were eliminated from the analysis.
Figure~\ref{fig:StelPhotKKS3}a shows a bright part of the galactic CMD
for stars within the rectangular area of $\sim\!2000\div2000$ pixels  
around the centre of the GC, which is equal to the projected area of $\sim\!1\times1$~kpc.
We used the distance modulus and extinction determined by Karachentsev et al. (2015b) 
(Table~\ref{tab:prop_gal} in this paper).
Green dots represent stars within a radius of $\sim$47~pc (4.6$\arcsec$) from the centre of the GC.
This is the area, where the azimuthally averaged surface brightness of the GC confidently exceeds 
the sky level around the cluster (Fig.~\ref{figKingKKS3}).
Four isochrones by Bertelli et al. (2008) are overplotted:  
$\rm log(Age)=10.10, Y=0.30, Z=0.0004$ (green), $\rm log(Age)=10.10, Y=0.30, Z=0.001$ (red),
and two additional isochrones of higher metallicity and younger age.
Only the bright parts of the RGB and AGB sequences are shown.
Fig.~\ref{fig:StelPhotKKS3}b illustrates X and Y HST frame positions of the stars analysed in 
Fig.~\ref{fig:StelPhotKKS3}a.
The central $1.6\arcsec \times1.8\arcsec$ region of the GC is empty of stars due to strong crowding effects. 
Another large empty region is located $\sim150$ pixels to the right from the GC in KKs\,3. 
This is the place of a bright foreground star projection.

\begin{figure}
\hspace{0.25cm}
\includegraphics[angle=-90,scale=0.305]{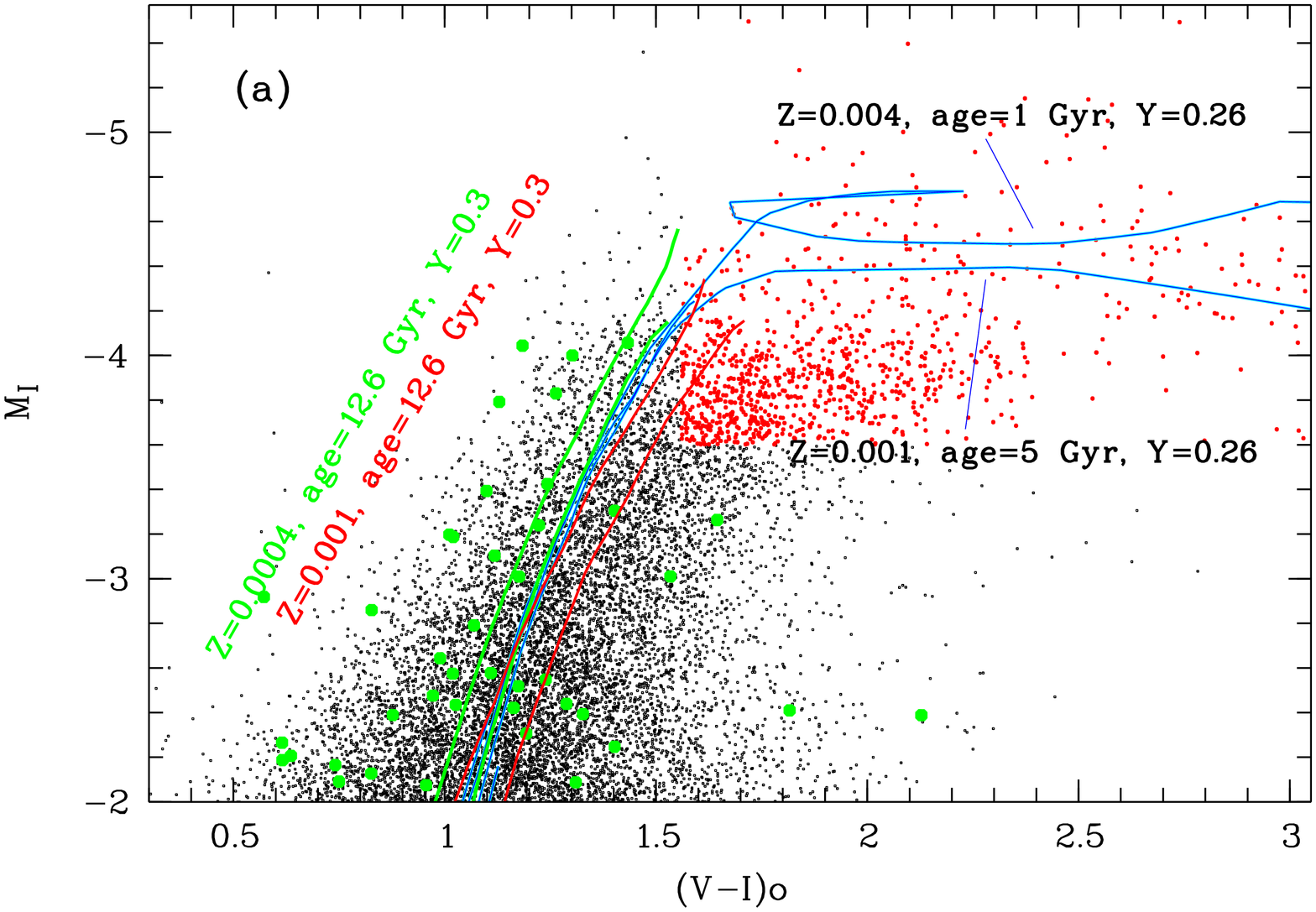}
\includegraphics[angle=-90,scale=0.32]{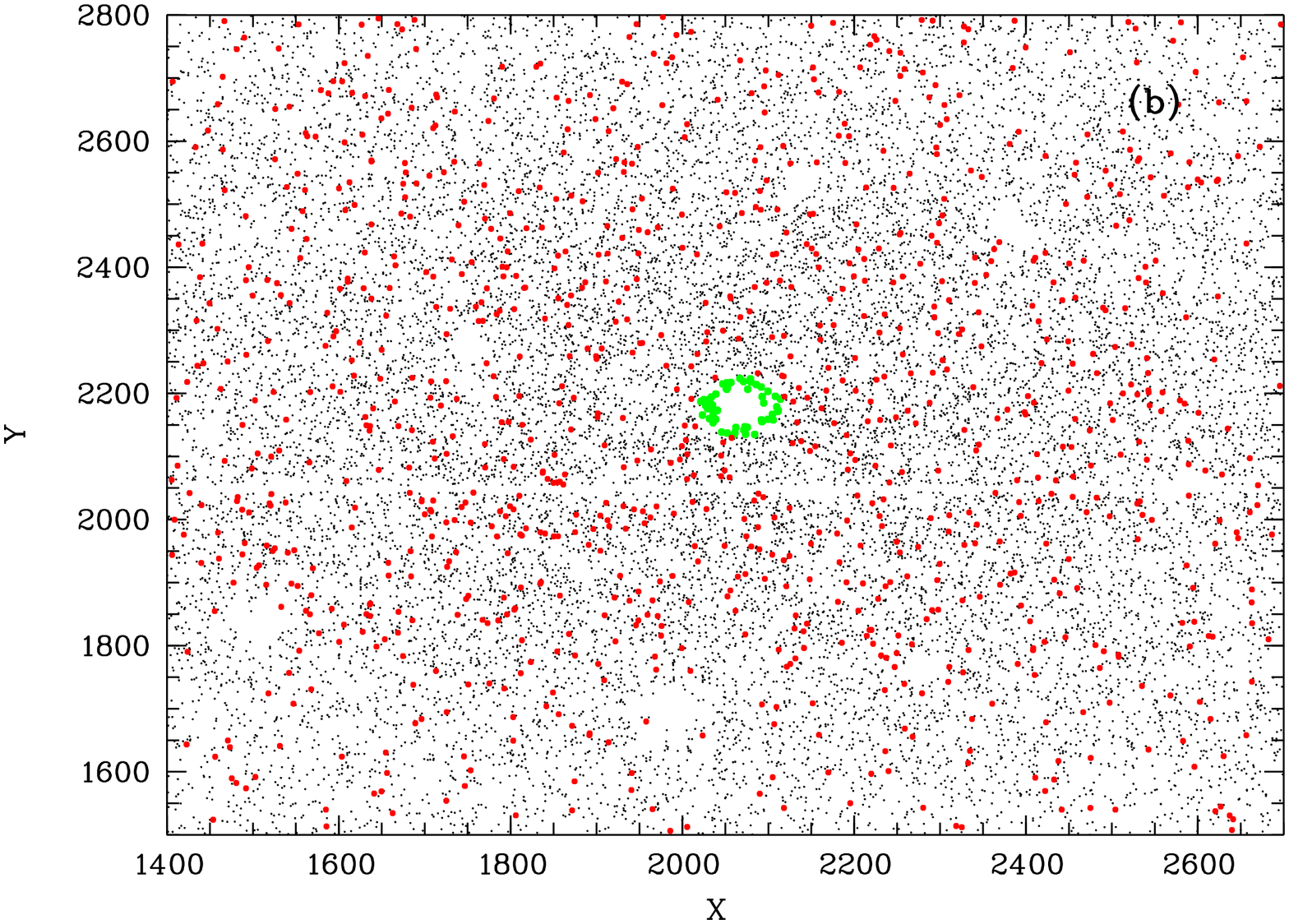}
\caption{The same as in Fig.~\ref{fig:StelPhotKKS3}, but for ESO\,269-66.}
\label{fig:StelPhotE269_66}%
\end{figure}

Similar analysis for stars in ESO\,269-66 is demonstrated in Fig.~\ref{fig:StelPhotE269_66}.
The magnitude, colour, and area selection criteria are the same as for KKs\,3.
We use stellar photometry data from Sharina et al. (2008) obtained using the DOLPHOT package (Dolphin 2002).
ESO\,269-66 is brighter than KKs\,3 and almost twice more distant. 
There are much more RGB and AGB stars in ESO\,269-66 than in KKs\,3 (Fig.~\ref{fig:StelPhotE269_66}a).
Photometry of only 54 stars was possible 
within a radius of $\sim$47~pc (2.5$\arcsec$) from the centre of the GC in ESO\,269-66 
(green dots in Fig.~\ref{fig:StelPhotE269_66}) 
due to strong crowding effects in the dense central $2.8\arcsec \times3.0\arcsec$ region of the GC.
Average photometric uncertainties at the level of TRGB $I=24.04$, $V-I\sim1.6$ are: $\sigma_I\sim 0.1$~mag,
$\sigma_V\sim 0.15$~mag (Karachentsev et al. 2007). 

Figs~\ref{fig:StelPhotKKS3}a and \ref{fig:StelPhotE269_66}a demonstrate that
the stars within the boundaries of the GCs in both dSphs have a bluer colour on average (lower metallicity)
than the rest of the upper RGB stars in the galaxies. Dispersion of stellar colours 
(metallicities) at $\rm M_I=-3.5\pm0.1$~mag in KKs\,3 is lower [$\sigma(V-I)=0.08$] than that in ESO\,269-66 
[$\sigma(V-I)=0.19$] (Sharina et al. 2008).

Figs~\ref{fig:StelPhotKKS3} and \ref{fig:StelPhotE269_66} show that stellar photometric data 
for the regions within a radius of $\sim$47~pc 
around the centres of the GCs in KKs\,3 and ESO\,269-66 do not contain bright red stars (red dots) with
magnitudes and colours in the range: $\rm M_I\le-3.6$~mag and $\rm (V-I)_0\ge1.56$~mag.
Probable contribution of bright intermediate-age metal-rich stars
in the central crowded regions of the clusters is 1--2 and 3--6 objects for KKs\,3 and ESO\,269-66, respectively.

Our spectroscopic analysis and the data presented in this section indicate
that intermediate-age and high-metallicity stellar populations do not 
contaminate significantly the integrated light of the studied GCs.


\clearpage
\section{Measuring of Lick indices in the spectra of GCs in KKs\,3 and E269-66}
\label{sec_lick}
\begin{figure*}
\centering
\begin{tabular}{p{0.49\textwidth}p{0.49\textwidth}}
\hspace{0.5cm}
\includegraphics[angle=-90,scale=0.37]{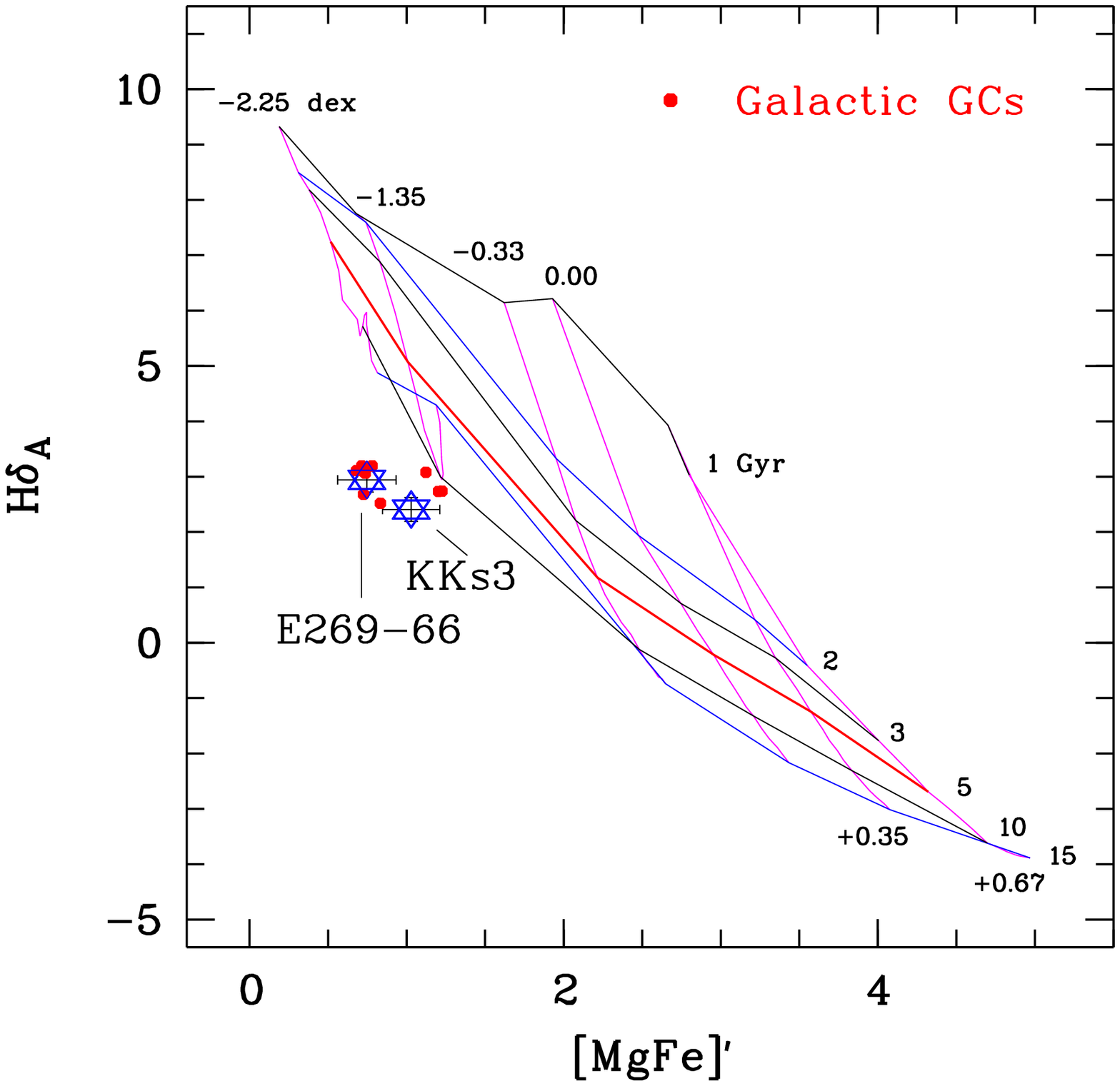} &
\hspace{0.5cm}
\includegraphics[angle=-90,scale=0.37]{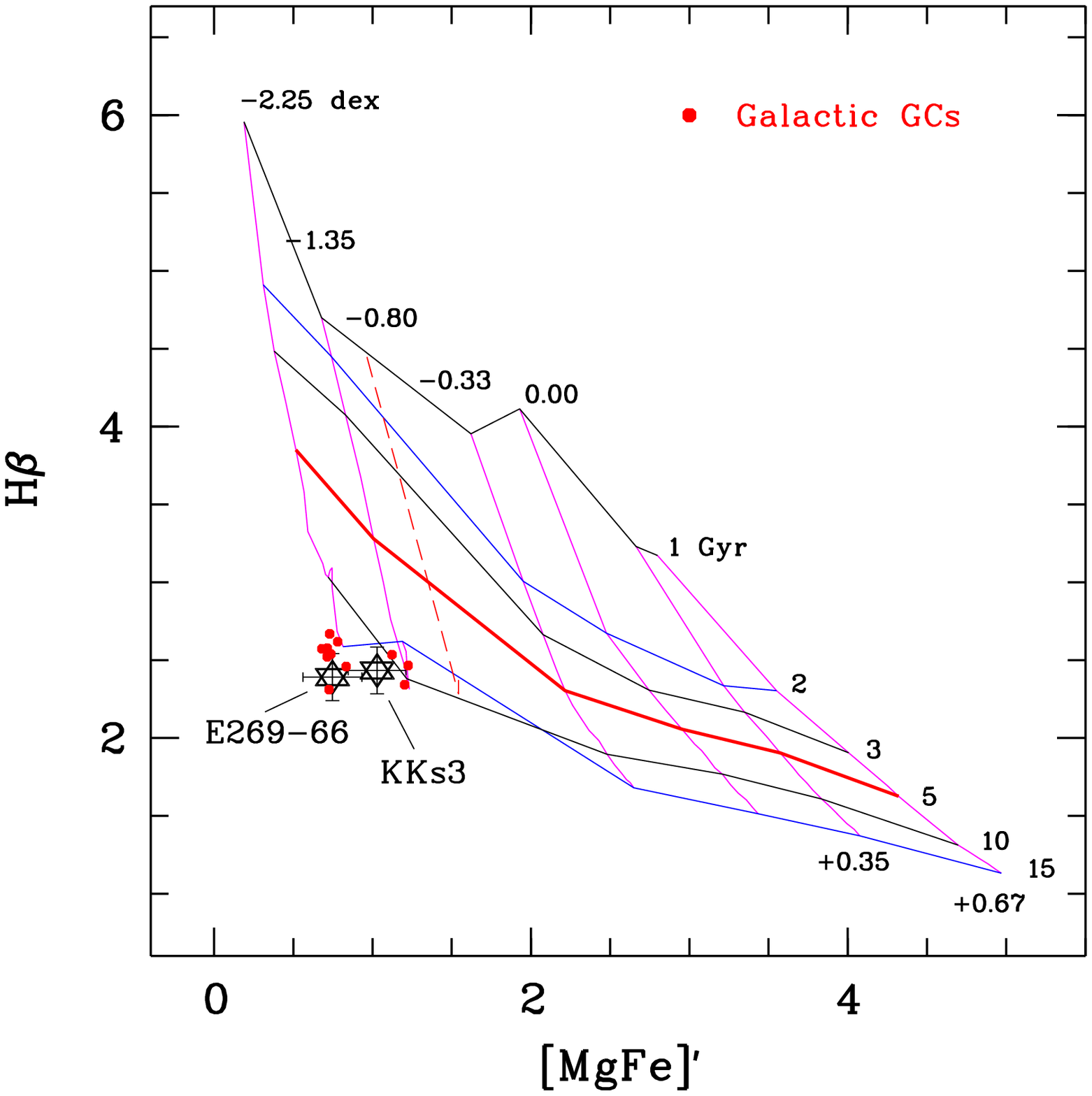} \\
\includegraphics[angle=-90,scale=0.37]{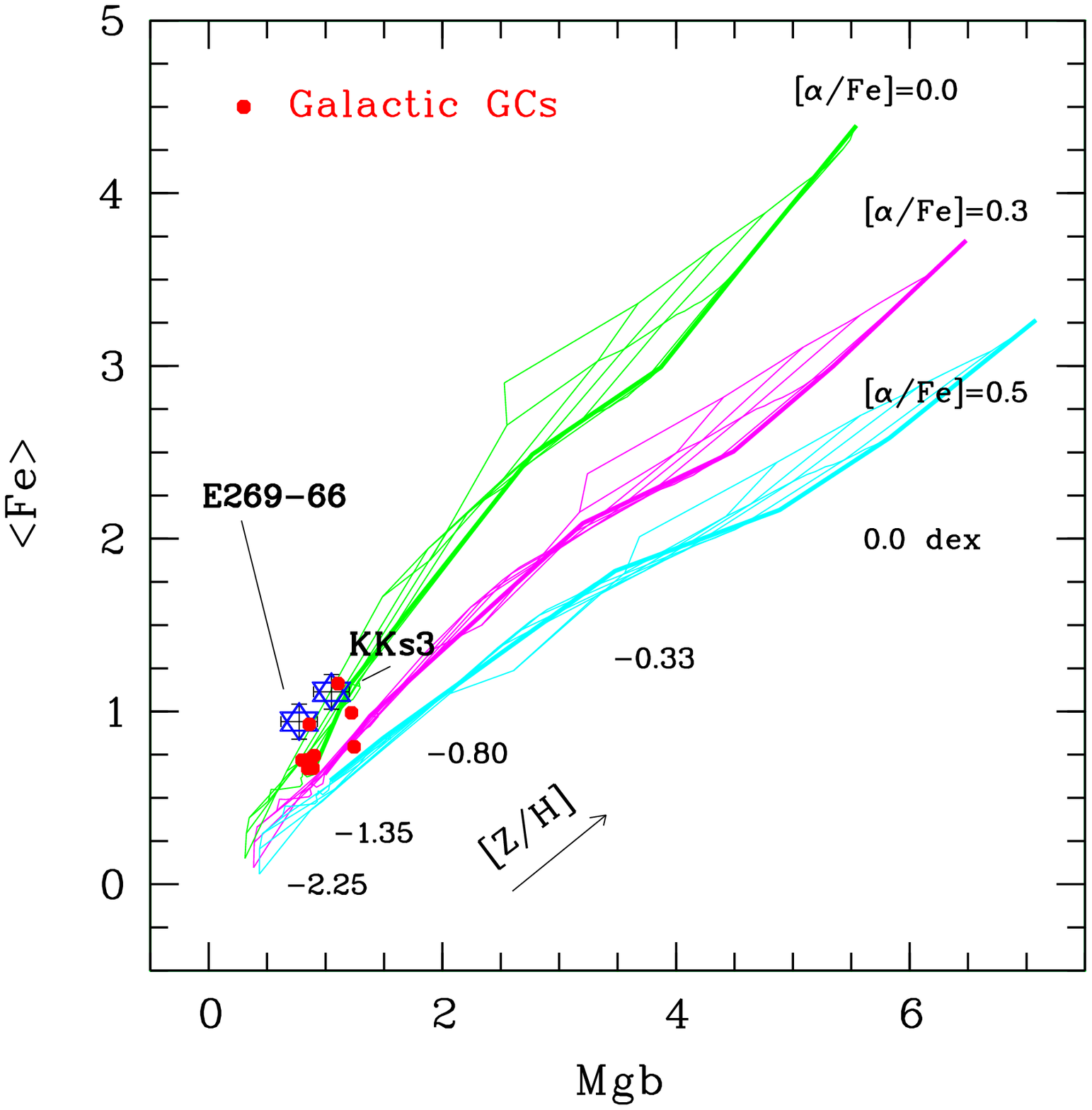} &
\includegraphics[angle=-90,scale=0.37]{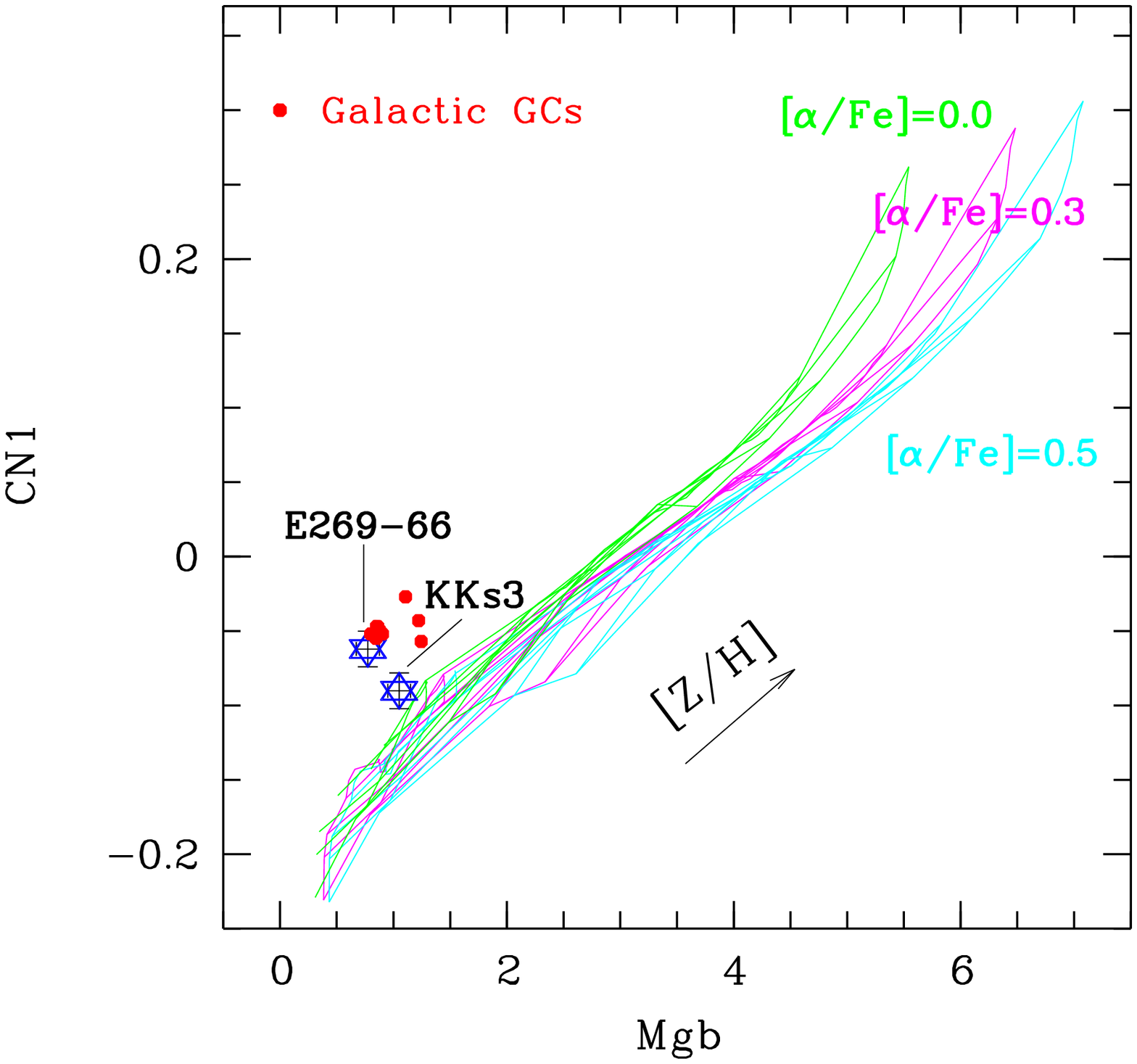} \\
\hspace{0.3cm}
\includegraphics[angle=-90,scale=0.37]{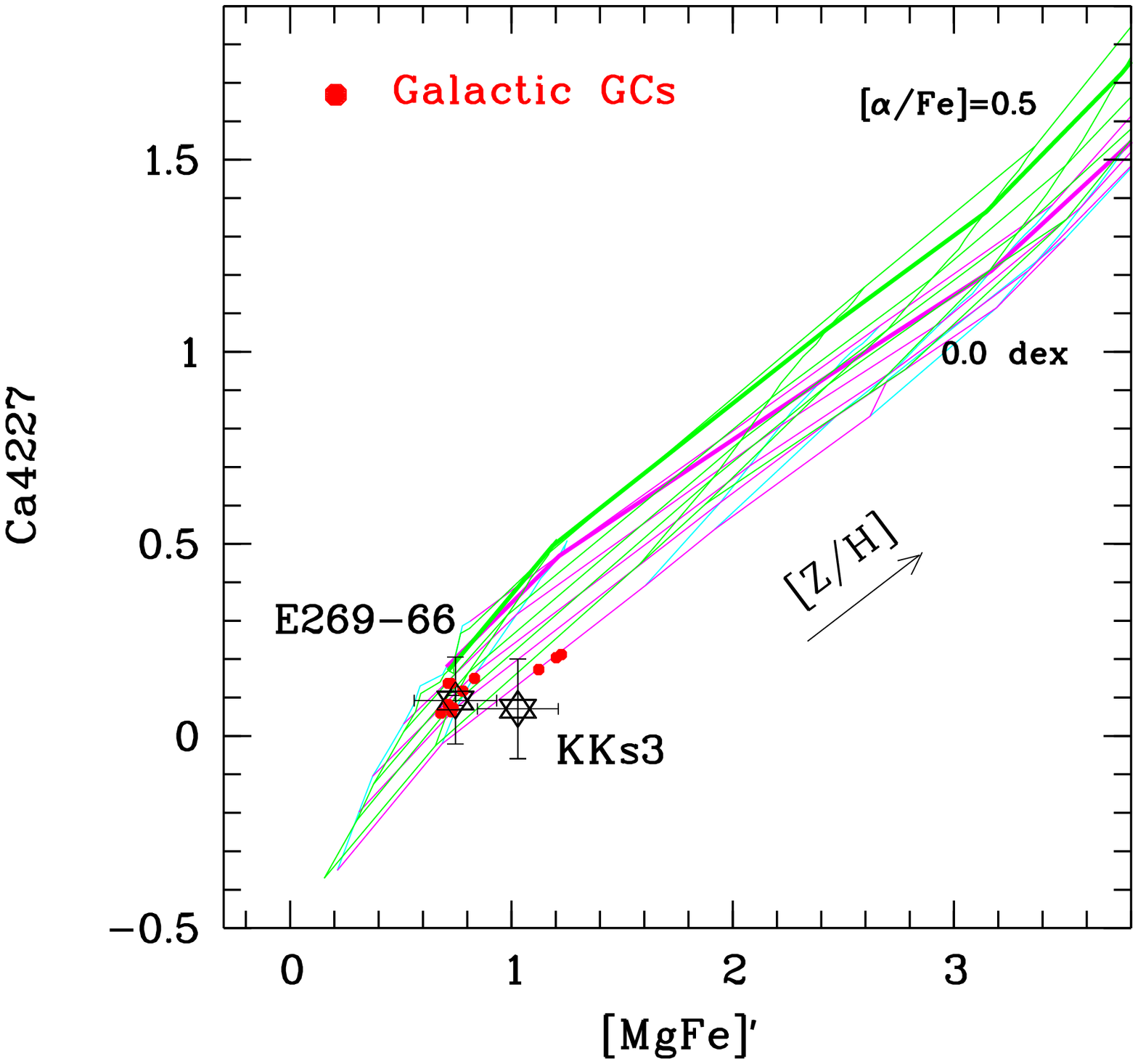} &
\hspace{0.3cm}
\includegraphics[angle=-90,scale=0.37]{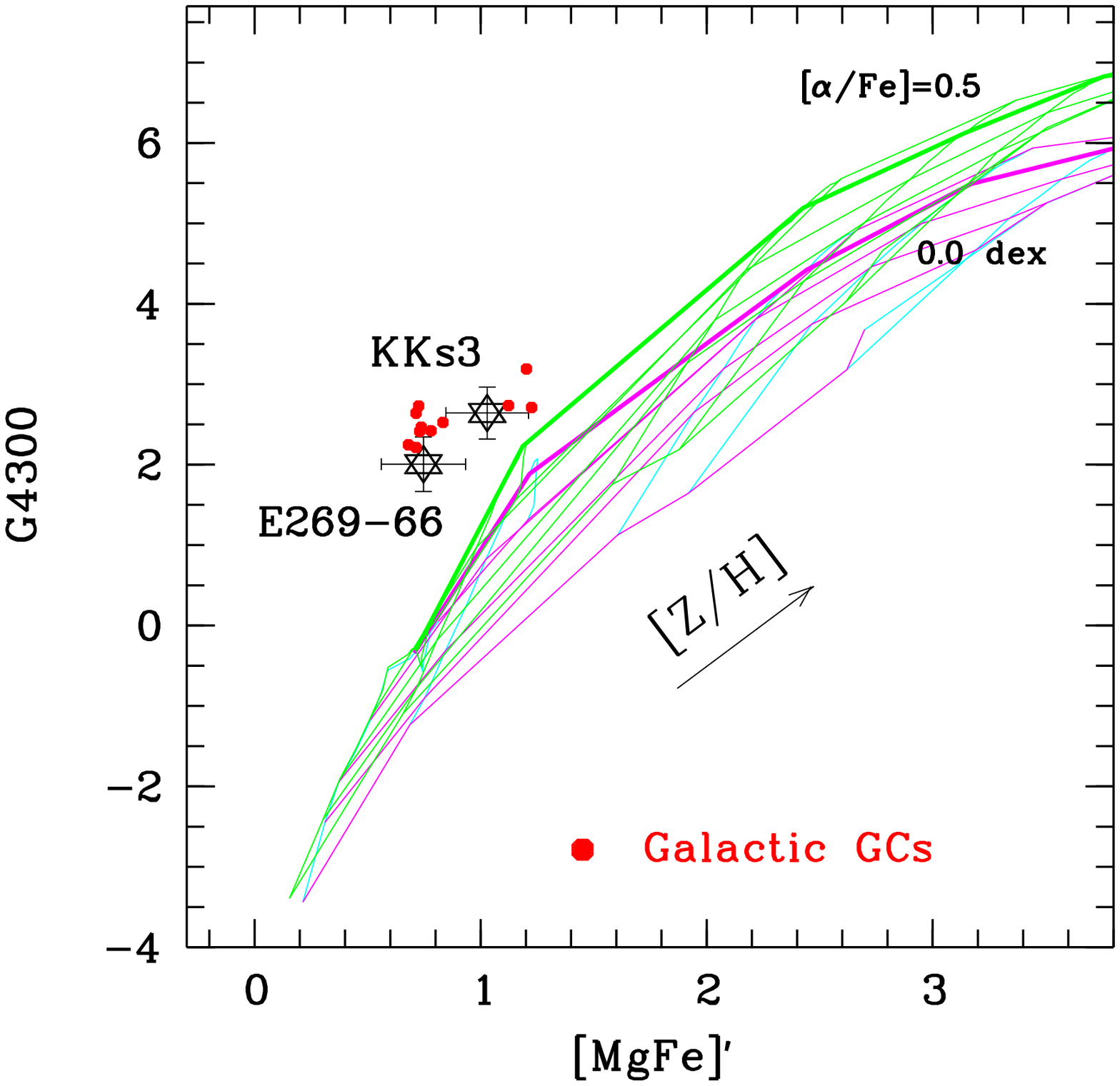} \\
\end{tabular}
\caption{Lick-index diagnostic plots for GCs in KKs\,3 and E269-66 and for Galactic GCs 
 (Schiavon et al. 2012) in two metallicity groups: 1) $\feh \sim -1.6$~dex 
 (NGC\,1904, 3201, 5286, 5946, 5986, 6254, 6333, 6752, 7089) and
2) $\feh \sim -1.3$~dex (NGC\,5904, 5946, 6218, 6235). SSP models by 
Thomas et a. (2003, 2004) are overplotted.}
\label{lick}%
\end{figure*}

Absorption-line indices in the so-called  Lick system (Burstein et al. 1984, Worthey et al. 1994, Worthey 1994,
 Worthey \& Ottaviani 1997, Trager et al. 1998) offer a useful tool 
 for disentangling age and metallicity effects on medium-resolution spectra of GCs and galaxies.
In this section, we present the
Lick indices for GCs in KKs\,3 and E269-66 measured using the flux-calibrated 
medium-resolution spectra (Section~\ref{observations}), the procedure described by Sharina et al. (2008),
and the calibrations into the Lick standard system by 
Katkov, Kniazev \& Sil'chenko (2015). Tables~\ref{lickind1} and \ref{lickind2} present the indices and their
 measurement errors.
Total errors of the Lick indices include the index measurement errors
and the errors of transformation into the Lick standard system (including random and systematic effects). 
If the signal-to-noise ratio in the studied spectra is $\rm S/N \ge 100$, the typical measurement errors 
are smaller than the errors of transformation into the Lick standard system (e.g. Schiavon et al. 2012).
In the following we list typical total errors at $\rm S/N\sim100$ for the indices presented in 
Fig.~\ref{lick}:
  $\rm \sigma (H \delta_A) \sim 0.7$\AA\ , $\rm \sigma (H \beta) \sim 0.2$\AA\ , 
  $\rm \sigma(<Fe>=(Fe5270+Fe5335)/2) \sim 0.25$\AA~,
$\rm \sigma (\mgfe=\sqrt{Mgb (0.72 Fe5270 + 0.28 Fe5335)})\sim 0.25$\AA~,  $\rm \sigma Mgb \sim 0.15$\AA~,
$\rm \sigma CN1 \sim 0.025$~mag, $\rm \sigma Ca4227\sim 0.2$\AA, and $\rm \sigma G4300\sim0.35$\AA.
These errors do not include 
the errors of simple stellar population models which are large at low metallicities (Thomas et al. 2011).

In Fig.~\ref{lick} we compare the indices for GCs in KKs\,3 and E269-66 and the corresponding indices
for several Galactic GCs of similar metallicity from Schiavon et al. (2012) and  Schiavon et al. (2005).
Galactic GCs in two metallicity groups are shown: 1) $\feh \sim -1.6$~dex (NGC\,1904, 3201, 5986, 6254, 6333, 6752, and 7089) 
and 2) $\feh \sim -1.3$~dex (NGC\,5904, 5946, 6218, and 6235).
The $\alpha$-enhanced simple stellar population (SSP) models by Thomas et al. (2003, 2004) are over-plotted.
The upper two plots ($\rm H \delta_A$ vs. \mgfe 
and $\rm H \beta$ vs. \mgfe) serve to disentangle the effects of age and metallicity on integrated spectra of GCs. 
Two groups of Galactic GCs with
$\feh \sim -1.6$~dex and $\feh \sim -1.3$~dex are separated well and it is seen that the \mgfe data for the 
GCs in dSphs are approximately in between the two groups of Galactic GCs. 
The GC in KKs\,3 is slightly more metal-rich, and/or Mg-rich than the GC in ESO\,269-66.
The differences in the Balmer line indices are mainly caused by age
and a number of hot HB stars contributing to the integrated spectra (e.g. Schiavon et al. 2004).
It is surprising for us that the hydrogen line indices $\rm H \delta_A$ and $\rm H \beta$ are also 
almost identical for seven of nine Galactic GCs despite of their different HB morphologies (Harris, 1996).
The influence of light elements on the spectra and the measured indices are seen in the lower four panels
of Fig.~\ref{lick}. It is seen that the light-element content of the GCs in 
dSphs is close to that of the lower-metallicity group of Galactic GCs.
CN1, Ca4227, and Mgb indices are almost identical for seven Galactic GCs with $\feh \sim -1.6$~dex. 
GC in KKs\,3 has roughly the same Ca and C abundance as the GC in ESO\,269-66 and more 
abundant in nitrogen. All these qualitative conclusions agree with the results of our full spectrum fitting 
analysis using stellar atmosphere models (Sections~\ref{method} and \ref{results}).
\begin{table*}
\caption{Lick indices ($\lambda \le 4531$\AA\ ) (first line)
measured in the spectra of nuclear GCs in KKs\,3 and ESO\,269-66
with the corresponding uncertainties (second line indicated by the "$\pm$" sign).}
\label{lickind1}
\begin{tabular}{llrrrrrrrrr} \\
\hline \hline
ID            & H$\delta_{\rm A}$ & H$\gamma_{\rm A}$& H$\delta_{\rm F}$ &  H$\gamma_{\rm F}$ & CN$_1$  & CN$_2$   & Ca4227 & G4300  & Fe4383 & Ca4455    \\  
               &(\AA)             & (\AA)          &  (\AA)              &  (\AA)             & (mag)    & (mag)   & (\AA)  & (\AA)  & (\AA)  & (\AA)     \\
 \hline
  GC in KKs\,3       & 2.38         & 0.33           & 1.60                 & 1.49              & -0.080  & -0.064   & 0.06   & 2.64   & 0.35   &  0.14     \\
 \hskip 70pt $\pm$ & 0.10         & 0.12           &  0.12                &  0.12             & 0.006   & 0.006    & 0.05   &  0.11  & 0.12   &  0.08     \\
GC in E269-66      & 2.94         & 0.63           & 1.99                 & 1.53              & -0.062  & -0.060   & 0.09   &  2.01  & 0.23   &  0.31     \\
 \hskip 70pt $\pm$ & 0.08         & 0.09           &  0.09                & 0.08              & 0.006   & 0.005    & 0.04   &  0.12  &  0.07  &  0.06     \\
\hline  \hline
\end{tabular}
\end{table*}
\begin{table*}
\caption{Lick indices ($\lambda \ge 4531$\AA\ ) (first line)
measured in the spectra of nuclear GCs in KKs\,3 and ESO\,269-66
with the corresponding uncertainties (second line indicated by the "$\pm$" sign).}
\label{lickind2}
\begin{tabular}{llrrrrrrrrr} \\
\hline \hline
ID                 & Fe4531 & Fe4668 & H$\beta$ & Fe5015 & Mg$_1$   & Mg$_2$     & Mg$b$     & Fe5270   & Fe5335    & Fe5406 \\
                   & (\AA) & (\AA)  &  (\AA)   & (\AA)  & (mag)     & (mag)      & (\AA)     & (\AA)    & (\AA)     & (\AA)  \\
 \hline
 GC in KKs\,3        &  0.99  & 0.25  & 2.42     & 1.43   & 0.037     & 0.067      & 1.03      & 0.78     & 1.46      & 0.50    \\
 \hskip 70pt $\pm$ &  0.09  &  0.17 & 0.06     & 0.15   &  0.003    &  0.003     &  0.04     & 0.06     & 0.08      & 0.05   \\
GC in E269-66      & 1.78   & 0.41  & 2.34     & 1.30  & 0.021      & 0.075      & 0.78      & 0.94     & 1.04      & 0.73    \\
 \hskip 70pt $\pm$ & 0.07   &  0.15 & 0.05     & 0.09  & 0.003      & 0.003      & 0.05      & 0.11     & 0.07      & 0.04     \\
\hline  \hline
\end{tabular}
\end{table*}

\bsp

\label{lastpage}

\end{document}